\documentclass[conference]{IEEEtran}
\IEEEoverridecommandlockouts
\usepackage{cite}
\usepackage{amsmath,amssymb,amsfonts}
\usepackage{adjustbox}
\usepackage{graphicx}
\usepackage{textcomp}
\usepackage{xcolor}
\usepackage{subcaption}
\usepackage{tcolorbox}
\usepackage{algorithm}
\usepackage{algorithmic}
\usepackage{paralist}
\usepackage{enumitem}
\usepackage{makecell}
\usepackage{xspace}
\usepackage{nicematrix}
\usepackage{booktabs}
\usepackage{tabularx}
\usepackage{lipsum}
\usepackage{multirow}
\usepackage[colorlinks=false,linkcolor=black,urlcolor=blue,citecolor=black]{hyperref}
\newcommand{\myauthornote}[3]{}

\newcommand{\model}{\textsc{ChatDANCE}\xspace}

\newcommand{\Fig}{Figure\xspace}

\newcommand{\Table}{Table\xspace}
\newcommand{\Sec}{Section\xspace}
\newcommand{\Eq}{Equation\xspace}

\usepackage{listings}
\def\BibTeX{{\rm B\kern-.05em{\sc i\kern-.025em b}\kern-.08em
    T\kern-.1667em\lower.7ex\hbox{E}\kern-.125emX}}

\begin{document}

\title{You Augment Me: Exploring ChatGPT-based Data Augmentation for Semantic Code Search}

\newcommand\notedauthorfootnote[1]{%
  \begingroup
  \renewcommand\thefootnote{}\footnote{\textsuperscript{\dag}#1}%
  \addtocounter{footnote}{-1}%
  \endgroup
}
\author{
Yanlin Wang\textsuperscript{a}
Lianghong Guo\textsuperscript{b}
Ensheng Shi\textsuperscript{c,\dag}
Wenqing	Chen\textsuperscript{a}
Jiachi	Chen\textsuperscript{a}\\
Wanjun	Zhong\textsuperscript{a}
Menghan	Wang\textsuperscript{d}
Hui	Li\textsuperscript{e}
Hongyu	Zhang\textsuperscript{f}
Ziyu	Lyu\textsuperscript{g}
Zibin	Zheng\textsuperscript{a}

\\
\textsuperscript{a}Sun Yat-sen University \quad
\textsuperscript{b}Beijing University of Posts and Telecommunications\\
\textsuperscript{c}Xi'an Jiaotong University   \quad
\textsuperscript{d}eBay Inc. \quad
\textsuperscript{e}Xiamen University \quad

\textsuperscript{f}Chongqing University \\
\textsuperscript{g}Shenzhen Institute of Advanced Technology, Chinese Academy of Sciences \quad
\\{ \{wangylin36, chenwq95, chenjch86, zhongwj25, zhzibin\}mail.sysu.edu.cn }\\
{ glhxystxdy123@gmail.com, 
   s1530129650@stu.xjtu.edu.cn }\\
{ wangmengh@zju.edu.cn,  hui@xmu.edu.cn }\\
{ hyzhang@cqu.edu.cn, zy.lv@siat.ac.cn } 
}

\maketitle

\begin{abstract}

Code search plays a crucial role in software development, enabling developers to retrieve and reuse code using natural language queries. While the performance of code search models improves with an increase in high-quality data, obtaining such data can be challenging and expensive. Recently, large language models (LLMs) such as ChatGPT have made remarkable progress in both natural and programming language understanding and generation, offering user-friendly interaction via simple prompts. Inspired by these advancements, we propose a novel approach ChatDANCE, which utilizes high-quality and diverse augmented data generated by a large language model and leverages a filtering mechanism to eliminate low-quality augmentations. Specifically, we first propose a set of ChatGPT prompting rules that are specifically designed for source code and queries. Then, we leverage ChatGPT to rewrite code and queries based on the according prompts and then propose a filtering mechanism which trains a cross-encoder from the backbone model UniXcoder to filter out code and query pairs with low matching scores. Finally, we re-train the backbone model using the obtained high-quality augmented data. Experimental results show that ChatDANCE achieves state-of-the-art performance, improving the best baseline by 13.2\% (R@1) and 7\% (MRR). Surprisingly, we find that this augment-filter-retrain strategy enables the backbone model (UniXcoder) to self-grow. Moreover, extensive experiments show the effectiveness of each component and ChatDANCE has stable performance under different hyperparameter settings. In addition, we conduct qualitative and quantitative analyses to investigate why ChatDANCE works well and find that it learns a more uniform distribution of representations and effectively aligns the code and query spaces. We have made the code and data anonymously available at \url{https://anonymous.4open.science/r/ChatDANCE}. \notedauthorfootnote{Ensheng Shi is the corresponding authors.}
\end{abstract}

\begin{IEEEkeywords}
Code Search, Data Augmentation, ChatGPT
\end{IEEEkeywords}

\section{Introduction}

With the rapid growth of open-source code repositories on platforms like GitHub~\cite{github}, code search has become crucial in software engineering. This task aims to find the code snippet in a repository that best matches users' intention, given a query written in natural language~\cite{allamanis2018survey}. Code search enables developers to find and reuse relevant code snippets in software development and maintenance~\cite{singer2010examination,nie2016query}.

Early studies~\cite{mcmillan2011portfolio,lu2015query,lv2015codehow,linstead2009sourcerer} in code search often rely on traditional information retrieval (IR) techniques, such as matching keywords based on lexical information of code snippets. With the popularity of deep learning, neural code search models~\cite{gu2018deep, cambronero2019deep, ling2021deep, shuai2020improving,du2021single,li2020learning,ye2020leveraging,mamulcs,zhu2020ocor,wan2019multi,haldar2020multi,gu2021multimodal,ling2020adaptive,sun2022code,shi2022better} begin to emerge. For example, DeepCS~\cite{gu2018deep} utilizes neural models to encode queries and codes into a shared vector space and measures similarity using vector distance. Later, pre-trained models~\cite{wang2021codet5,feng2020codebert,guo2020graphcodebert,guo2022unixcoder,shi2023cocosoda,ahmad2021unified,bui2021self,jain2021contrastive} emerge and surpass the conventional neural models in code search. These models better understand source code and natural language by pre-training on vast amounts of code and natural language data. Finetuning such models can achieve excellent results on downstream tasks such as code search. For example, UniXcoder~\cite{guo2022unixcoder} is a unified cross-modal pre-trained model for programming languages that utilize mask attention matrices with prefix adapters to control the model's behavior and leverages cross-modal contents like AST and code comment to enhance code representation. By finetuning, UniXcoder significantly improves most downstream tasks, such as code search and summarization.

Despite the significant advantages of deep learning, two main bottlenecks prevent neural models from achieving high performance: 1) the lack of high-quality labeled training data and 2) the difference in data distribution between the training and testing datasets~\cite{dong2023boosting}. To overcome these challenges, a straightforward solution is to increase the size and diversity of the training data by data augmentation. 

\begin{figure*}[t]
    \centering
    \includegraphics[width=\textwidth, height=0.4\textheight]{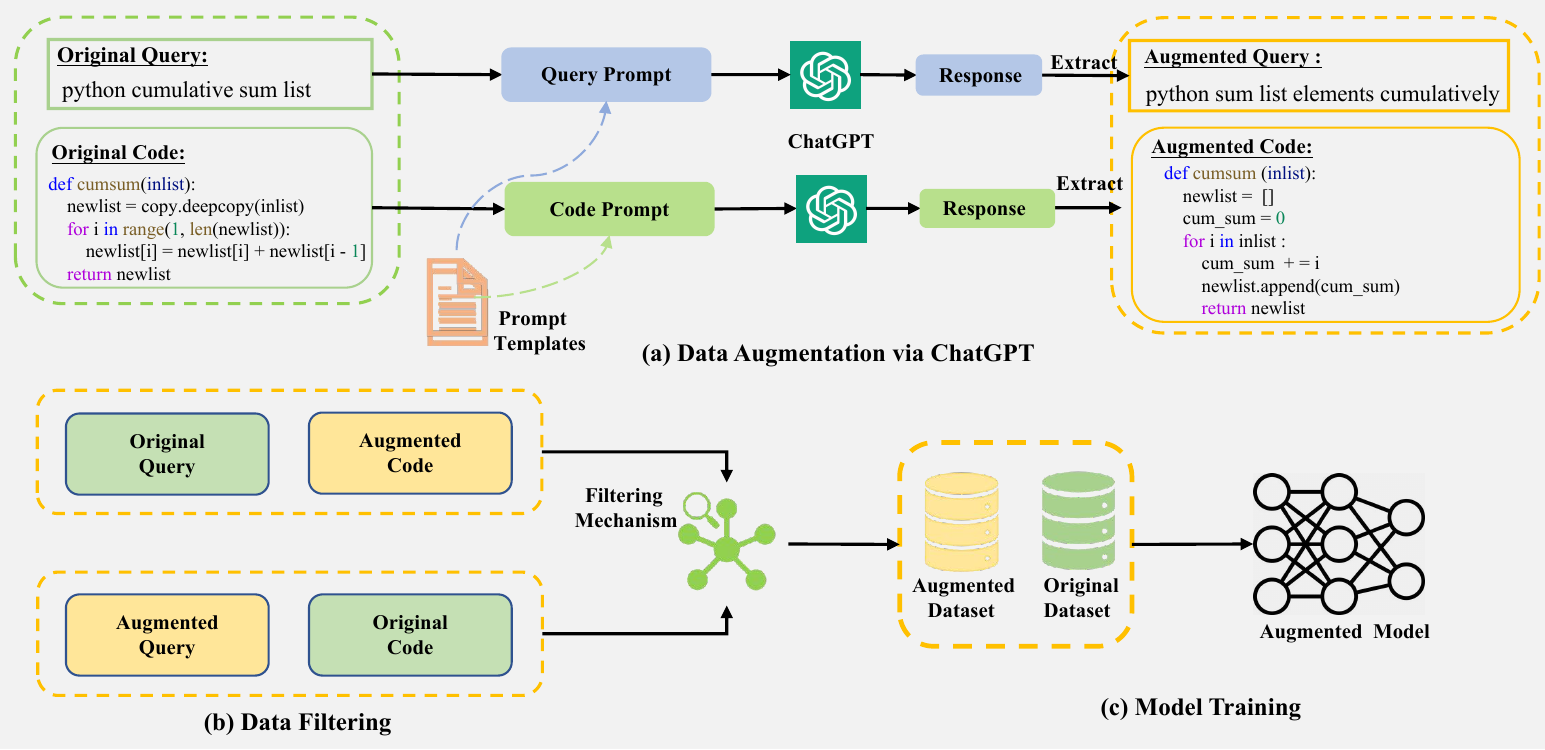}
    \caption{An overview of \model.}
    \label{fig:overview}
\end{figure*}

Recently, large language models (LLMs) such as ChatGPT have made remarkable progress in both natural and programming language understanding and generation, offering user-friendly interaction via simple prompts. Inspired by these advancements, we propose \textbf{\model}, \textbf{Chat}GPT-based \textbf{D}ata \textbf{A}ugme\textbf{N}tation for \textbf{C}ode s\textbf{E}arch). Our method can exploit LLMs such as ChatGPT to generate a large number of high-quality and diverse data using two core components: ChatGPT-based data augmentation and model-based data filtering. First, we use the concept of data augmentation~\cite{van2001art} and request the LLM to rewrite data by making semantic-similar modifications to the original data, which can effectively preserve most of the semantics of the generated data. To efficiently interact with the model and perform data augmentation, we designed two prompt templates for queries and codes, respectively. These templates contain essential information for the rewriting task, such as task definition and critical additional information. Moreover, our method allows users to inject specific prior knowledge to guide the LLM in data augmentation. For instance, in code augmentation, we designed five rewriting techniques to guide the model to generate new data according to specific patterns. Second, to further improve the quality of the generated data, we trained a filtering model on the original dataset to score and filter the augmented samples. The filtering model evaluates the quality of the generated data and discards low-quality data. With our proposed \model method, we can generate a large number of diverse and high-quality data, which can be used to enhance the generalization ability and performance of models in downstream tasks such as code search.

The overall framework of \model is presented in \Fig~\ref{fig:overview}. Our approach consists of three stages: (a) data augmentation via ChatGPT, (b) data filtering, and (c) model training. In part (a) of \Fig~\ref{fig:overview}, we demonstrate how we achieve data augmentation using ChatGPT. Given a query-code pair, we first construct query and code augmentation prompts using different prompt templates for the query and code, respectively. The prompt template is crucial for effectively interacting with ChatGPT. It includes the definition of the data rewriting task, important additional information such as the number of augmented data, and prior knowledge for completing the task. We then request ChatGPT to rewrite the original data based on the prompt information, generating augmented data. Finally, we extract the augmented data from ChatGPT's response using regular expressions.

In the data filtering stage, we first train a filtering model to remove low-quality augmented data to further improve data quality. Specifically, we train a model based on the cross-encoder architecture, which can directly score the matching degree of query-code pairs. We use the scores generated by the filtering model as the basis for filtering the data. Next, we use the filtering model to score the matching degree of the augmented query-code pairs. We filter out query-code pairs with scores below the filtering threshold and obtain high-quality data. By removing low-quality data, we can ensure that the generated data is of high quality and can effectively enhance the model's performance.

Finally, after filtering the augmented dataset, we combine it with the original dataset. Then, we use UniXcoder as our baseline model with a bi-encoder structure and fine-tune the model on the final dataset using contrastive loss to improve the model's performance.

We evaluate the effectiveness of our approach on the CoSQA dataset, which contains a large number of real-world queries. We apply our approach to the state-of-the-art model UniXcoder in the CoSQA dataset and compare our approach with semantic-preserving data augmentation methods: QRA and NatGen. We also conduct ablation studies to investigate the effectiveness of each component of our approach and explore the impact of different hyperparameters on our method. Finally, we conduct qualitative and quantitative analyses to investigate why our approach works. The results of the experiments demonstrate that: (1) Our approach can significantly improve the model's performance and outperform the baselines. (2) Each component of our approach contributes significantly to improving the model's performance. (3) Our method ensures stable performance for the model across various hyperparameter settings, including query filtering threshold ranging $\theta_{q}$ from 0.7 to 0.95, code filtering threshold $\theta_{c}$ ranging from 0.7 to 0.9, learning rate ranging from 1e-5 to 5e-5 and the average number of augmented samples greater than 5. (4) Compared to baselines, our method can effectively improve the alignment and uniformity of the representations learned by the model.

We summarize the contributions of this paper as follows:

\begin{itemize}
    \item We propose a new data augmentation approach for code search, which uses ChatGPT to generate a large number of high-quality data. We also introduce a prompt schema to improve the interaction with ChatGPT, allowing users to design their own prompts and provide the necessary information to complete the task effectively.

    \item We propose a cross-encoder-based data filtering mechanism that scores code-query pairs and filters low-quality pairs. This approach can be applied to data collection and augmentation, resulting in improved data quality.

    \item We conduct extensive experiments to evaluate the effectiveness of our method on the CoSQA dataset. The experimental results show that our method can significantly outperform baselines. Furthermore, our method can effectively improve the alignment and uniformity of the learned representations and exhibits stable performance across a range of hyperparameters.
    
\end{itemize}

\section{Related Work}

\subsection{Code Search}
Code search is a crucial aspect of software development and maintenance~\cite{singer2010examination,nie2016query}. In general, models for code search can be classified into two categories: information retrieval (IR)-based models~\cite{mcmillan2011portfolio,lu2015query,lv2015codehow,linstead2009sourcerer} and deep learning-based models~\cite{gu2018deep,du2021single,shuai2020improving,bui2021self,cambronero2019deep,shi2022better,sun2022code,jain2021contrastive,ling2021deep,wan2019multi,haldar2020multi,li2020learning,ye2020leveraging,mamulcs,zhu2020ocor,gu2021multimodal,ling2020adaptive,wang2021codet5,feng2020codebert,guo2020graphcodebert,guo2022unixcoder,shi2023cocosoda,ahmad2021unified}. IR-based models often use keyword matching or text similarity to retrieve relevant code. In recent years, deep learning-based models have become the mainstream approach for code search and have achieved promising results. For instance, Gu et al.~\cite{gu2018deep} proposed the first neural code search model, CODEnn, which embeds queries and code into a shared vector space and calculates similarity by vector distance. Since then, various deep learning-based models have been developed, including sequence models~\cite{cambronero2019deep,wan2019multi,ye2020leveraging,haldar2020multi,shuai2020improving,sun2022code}, convolutional neural networks~\cite{li2020learning,zhu2020ocor,ling2020adaptive}, and graph neural networks~\cite{ling2021deep,wan2019multi}. Recently, pre-trained models~\cite{wang2021codet5,feng2020codebert,guo2020graphcodebert,guo2022unixcoder,shi2023cocosoda,ahmad2021unified,jain2021contrastive,bui2021self} have emerged as a powerful tool for code search, outperforming traditional neural models. Pre-trained models leverage massive amounts of programming language and natural language data to develop strong code understanding capabilities and achieve excellent performance on various code-related tasks, including code search. For example, Feng et al.~\cite{feng2020codebert} proposed CodeBERT, a bimodal pre-trained model that learns representations for both programming language and natural language. Guo et al. proposed GraphCodeBERT~\cite{guo2020graphcodebert}, which utilizes data flow to pre-train the model and improve the representation of code. UniXcoder~\cite{guo2022unixcoder}, a unified cross-modal pre-trained model for programming languages, uses mask attention matrices with prefix adapters to control the model's behavior and leverages cross-modal contents, such as abstract syntax trees and code comments, to enhance code representation. In this paper, we adopt UniXcoder as our baseline model, as it is the state-of-the-art model on the CoSQA dataset~\cite{huang2021cosqa}.

\subsection{Large Language Model and In-context Learning}
Large language models (LLMs) typically refer to language models with hundreds of billions or more parameters~\cite{zhao2023survey}. Trained on large amounts of text data, LLMs such GPT-3~\cite{brown2020language}, PaLM~\cite{chowdhery2022palm}, and LLaMA~\cite{touvron2023llama} demonstrate impressive performance on various downstream tasks such as machine translation, code generation, and more. As the model parameters and size of training data further increase, some emergent abilities of LLMs have been observed when the model size exceeds a certain level.

One of large language models' impressive emergent abilities is their in-context learning capability. The in-context learning ability is formally introduced in GPT-3~\cite{brown2020language}, which enables the model to generate the expected output for test instances by completing the input text's word sequence, given natural language instructions and/or task demonstrations~\cite{zhao2023survey}. Inspired by the concept of in-context learning, we introduce a novel data augmentation approach that leverages large language models such as ChatGPT. By providing task instructions to the model, we can prompt it to rewrite existing data and generate high-quality augmented data.

\subsection{Data Augmentation in Code Search}
Data augmentation is a common method used in code search to help models achieve better generalization ability and performance. For instance, Huang et al.~\cite{huang2021cosqa} propose Query-Rewritten Augmentation (QRA) to generate augmented queries by conducting small modifications on queries. The QRA performs query augmentation by three transformations such as (1) deleting a word randomly, (2) copying a word randomly, and (3) switching the position of two words randomly. In addition, Chakraborty et al.~\cite{chakraborty2022natgen} propose a code augmentation method called NatGen, which includes six semantics-preserving transformations: (1) Loop Transformation, (2) DeadCode Injection, (3) Operand Swap, (4) Block Swap, (5) Variable Renaming, (6) Confusing Code Insertion. This method chooses appropriate code transformations based on the AST structure to rewrite the code. Later, Jain et al.~\cite{jain2021contrastive} and Bui et al.~\cite{bui2021self} adopt similar code augmentation methods for pre-training their models. Recently, Shi et al.~\cite{shi2023cocosoda} proposed a non-semantics-similar method called Soft Data Augmentation (SoDa) to pre-train code search models by dynamically masking tokens in the data. Compared to the previous approach, ChatGPT Data Augmentation is semantics-similar and suitable for both queries and codes. Moreover, by exploiting the powerful generation ability of LLM, our method can generate data with better diversity and scalability.

\section{\model Framework}
\label{approach_sec}
This section introduces our straightforward and effective data augmentation framework via ChatGPT for augmenting training data on code search. The framework consists of three subsequent stages, \textit{the data augmentation stage}, \textit{the data filtering stage}, and \textit{the model training stage}. An overview of the framework, when applied to augment query-code pairs in the training dataset, is shown in \Fig~\ref{fig:overview}. 

In the first stage, we separately augment both query and code modalities to create augmented samples. For query augmentation, we request ChatGPT to rewrite an input query without changing its semantics. Then we pair the rewritten query with its original code to form a new augmented sample. For code augmentation, we request ChatGPT to rewrite the code with the guidance of the five given rewriting techniques, and then we pair the rewritten code with its original query to form a new augmented sample. 

Next, considering the existence of low-quality augmented samples that may introduce noise for model training, in the second stage, we trained a cross-encoder model to compute matching scores for augmented query-code pairs. The matching scores serve as the basis for filtering out sample pairs that do not meet a certain threshold. 

Finally, we train the model on the augmented dataset to improve its performance. In the following, we will provide a detailed explanation of the design of each stage.

\subsection{The Data Augmentation Stage}
\label{subsec_data_aug}

\subsubsection{Prompt Schema}
\label{sub_sec_prompt_schema}
Pre-trained on massive and unlabeled corpora, Large Language Models (LLMs) such as ChatGPT have demonstrated impressive emergent capabilities when subjected to model scaling. Instead of fine-tuning large models on specific tasks, complex problems such as machine translation and code generation can be solved simply by interacting with the LLMS using appropriate prompts. Therefore, constructing a good prompt is critical to effectively using a large language model. Inspired by~\textit{Natural Instruction}~\cite{mishra2022cross}, we develop a similar prompt schema to build our data augmentation request to ChatGPT. Below we present the ingredients of our schema:
\begin{itemize}
  \item \underline{Instruction} provides detailed content about the task, which often includes task input, task output, and approach to complete the task.
  \item \underline{Emphasis and Caution} provides important additional requirements to ensure the effective completion of the task.
  \item \underline{Prior Knowledge} provides ChatGPT with prior knowledge to efficiently accomplish the rewriting task.
  \item \underline{Task Input} provides the input content for the task. 
  \item \underline{Outputs Context} provides ChatGPT with the context of returning task outputs.
\end{itemize}

\subsubsection{Prompt Design}
Following the prompt schema in \Sec~\ref{sub_sec_prompt_schema}, We design different prompts based on the characteristics of query and code data, respectively. Next, we separately elaborate on the content of the query prompt and code prompt.

\textbf{\textit{Query Augmentation Prompt: }}The details of the query augmentation prompt is shown in \Table~\ref{tab:query_aug_table}. (1) \textit{Instruction}: We aim to generate augmented queries through data reformulation to increase the diversity of queries. Therefore, we request ChatGPT reformulate the query without changing its original semantics. (2) \textit{Emphasis}: We firstly specify to ChatGPT the desired quantity of generated queries. And then, we illustrate to ChatGPT using the CoSQA dataset example that queries are brief, ensuring that it generates concise queries. (3) \textit{Caution}: To further ensure the brevity of queries, we limit the length of the augmented query by \Eq~\ref{length_aug_query}. In the experiment, we set the $\alpha$ to 1.6.\begin{equation}
Length_{origianl}<= Length_{aug}<= \alpha * Length_{original}
\label{length_aug_query} \\
\end{equation} (4) \textit{Prior Knowledge}: Due to the brevity of queries, the task is relatively simple to complete without providing prior knowledge for query reformulation. So we do not provide prior knowledge here. (5) \textit{Task Input \& Output Context}: We provide the original query as the task input and provide the text ``Rewritten Queries'' as the context for ChatGPT to directly return results.
\begin{table}[H]
\caption{Structure of query augmentation prompt.}
\label{tab:query_aug_table}
\begin{tabular}{ll}
\toprule
\textbf{Component}                                 & \textbf{Content}                                   \\ \midrule
Instruction &
  \multicolumn{1}{l}{\begin{tabular}[c]{@{}l@{}}Given a query, your task is to reformulate the query \\ while ensuring that its semantics remain unchanged. 
    .\end{tabular}} \\ \hline
Emphasis &
  \multicolumn{1}{l}{\begin{tabular}[c]{@{}l@{}}You must generate (15) queries. Note that in real-\\life scenarios, users'queries are often brief. For \\example, the average length of queries in CoSQA \\ dataset is 6.6. So you must aim to generate concise \\ queries in this task. \end{tabular}} \\ \hline
Caution &
  \multicolumn{1}{l}{\begin{tabular}[c]{@{}l@{}}You must limit the length of each rewritten query \\ to between (query\_length) and 1.6 *(query\_length).\end{tabular}} \\ \hline
\multicolumn{1}{l}{Prior Knowledge} & \textbackslash{}                               \\ \hline
Task Input                                & Original Query: \textless{}Query\textgreater{} \\ \hline
Output Context                            & Rewritten Queries:                             \\\bottomrule
\end{tabular}
\end{table}

\textbf{\textit{Code Augmentation Prompt: }}The details of the code augmentation prompt are shown in \Table~\ref{tab:code_aug_table}. (1) \textit{Instruction}: Similar to the query augmentation prompt, we request ChatGPT to rewrite code without changing its functionality in code enhancement. In addition, we provide a rewriting technique to guide ChatGPT in rewriting the code. (2) \textit{Emphasis}: We firstly specify to ChatGPT the desired quantity of generated codes. And then, we ask ChatGPT to return codes following a given template so that the rewritten codes can be automatically extracted via regular expression. If the given rewriting technique is not suitable for rewriting, we allow ChatGPT to use different methods to rewrite the code to avoid returning an empty response. (3) \textit{Prior Knowledge}: Compared to query, code is usually longer and contains more rich information, which provides more space for code rewriting. Here we propose five rewriting techniques to guide ChatGPT in efficiently rewriting code based on different levels of information. The details are follows. 
\begin{compactitem}
\label{code_aug_prompt}
  \item \textbf{Rename the method without changing the function names it calls internally.}
  \item \textbf{Rewrite the code with more meaningful variable names.}
  \item \textbf{Use different library functions for the code snippet.}
  \item \textbf{Rewrite the code with the same semantics.}
  \item \textbf{Simplify the code by removing unnecessary statements or tokens.}
\end{compactitem}
(4) \textit{Task Input \& Output Context}: We provide the original code as the task input and provide the text ``Rewritten Codes'' as the context for ChatGPT to directly return results based on the given template.

\begin{table}[t]
\caption{Structure of code augmentation prompt.}
\label{tab:code_aug_table}
\begin{tabular}{ll}
\toprule
\textbf{Component}     & \textbf{Content}       \\ \midrule
Instruction &
  \multicolumn{1}{l}{\begin{tabular}[c]{@{}l@{}}Given a method-level code snippet, your job is to \\ rewrite the code snippet based on a given rewrit-\\ ing technique, while ensuring that the generated \\ code performs the same functionality as the origi-\\ nal code.\end{tabular}} \\ \hline
Emphasis &
  \multicolumn{1}{l}{\begin{tabular}[c]{@{}l@{}}You must generate (3) codes. And use ``` to wrap \\ each code based on this template : Code (number \\ such as 1)\textbackslash{}n```python\textbackslash{}n\textless{}returned code\textgreater{}\textbackslash{}n```. If \\ current rewriting technique is not suitable for the\\ original code, you can rewrite it using different \\ technique, while ensuring the generated code has \\ the same functionality as the original code. \end{tabular}} \\ \hline
\multicolumn{1}{l}{Prior Knowledge} & Rewriting Technique:\textless{}Rewriting Technique\textgreater{} \\ \hline
Task Input                                & Original Code: \textless{}Code\textgreater{}                     \\ \hline
Output Context                            & Rewritten Code:                                                 \\ \bottomrule
\end{tabular}
\end{table}

\subsubsection{Data Augmentation via ChatGPT}
\label{data_aug_via_chatgpt}
We use the ChatGPT API (gpt-3.5-turbo-0301) with default parameter settings to perform data augmentation. For query augmentation, we request ChatGPT to generate 15 augmented queries for each query based on the prompt template shown in \Table~\ref{tab:query_aug_table}. For code augmentation, we utilize five different rewriting techniques mentioned in \Sec~\ref{code_aug_prompt} to create five prompts based on the templates shown in \Table~\ref{tab:code_aug_table}. We generated 15 augmented codes per original code, using each of the five prompts to generate three new codes. After receiving a response from ChatGPT, we use regular expressions to extract the generated data. 

As an example, \Table~\ref{tab:query_examples} shows the five augmented queries generated by ChatGPT through the rewriting of the original query. We can see that ChatGPT has a good understanding of the semantics of the query and can produce high-quality and diverse queries that preserve the original semantics by replacing synonyms, changing syntax structures, and using other techniques. Overall, the generated queries are of high quality and exhibit good diversity.


\begin{table}[H]
\centering
\renewcommand\arraystretch{1.1}
\caption{Augmented query samples generated by ChatGPT query augmentation.}
\label{tab:query_examples}
\begin{NiceTabular}{ll}
\toprule
\textbf{Original Query}   & Math function for area of triangle python \\ \midrule
\textbf{Augmented Query 1} &Calculate triangle area in Python\\
\textbf{Augmented Query 2} &Triangle area formula in Python\\
\textbf{Augmented Query 3} &Triangle area algorithm in Python\\
\textbf{Augmented Query 4} &Python area calculation for triangle\\
\textbf{Augmented Query 5} &Formula to calculate triangle area in Python\\
\bottomrule
\end{NiceTabular}
\end{table}

\begin{figure*}[ht]
    \centering
    \hrule
    \vspace{0pt}
    \begin{subfigure}[t][0.12\textheight]{0.45\linewidth}
        \begin{minipage}[t][0.09\textheight]{\linewidth}
        \begin{lstlisting}[language=Python,frame=none,columns=fullflexible,numbers=left,numbersep=5pt,aboveskip=0pt,belowskip=0pt]
def get_tri_area(pts):
    a, b, c = pts[0], pts[1], pts[2]
    v1 = np.array(b) - np.array(a)
    v2 = np.array(c) - np.array(a)
    area_tri = abs(sp.linalg.norm(sp.cross(v1, v2)) / 2)
    return area_tri
        \end{lstlisting}
        \end{minipage}
        \caption{Original code}
    \end{subfigure}
    \hspace{0.02\linewidth}
    \vrule
    \hspace{0.02\linewidth}
    \begin{subfigure}[t][0.12\textheight]{0.45\linewidth}
        \begin{minipage}[t][0.09\textheight]{\linewidth}
        \begin{lstlisting}[language=Python,frame=none,columns=fullflexible,numbers=left,numbersep=5pt,aboveskip=0pt,belowskip=0pt,escapeinside=||]
def |\colorbox{orange!30}{calculate\_triangle\_area}|(pts):   
    a, b, c = pts[0], pts[1], pts[2]  
    v1 = np.array(b) - np.array(a)  
    v2 = np.array(c) - np.array(a)  
    area_tri = abs(sp.linalg.norm(sp.cross(v1, v2)) / 2) 
    return area_tri
        \end{lstlisting}
        \end{minipage}
        \caption{Rewrite method name}
        \label{name_rewrite}
    \end{subfigure}
    \hrule
    \begin{subfigure}[t][0.15\textheight]{0.45\linewidth}
        \begin{minipage}[t][0.12\textheight]{\linewidth}
        \begin{lstlisting}[language=Python,frame=none,columns=fullflexible,numbers=left,numbersep=5pt,aboveskip=0pt,belowskip=0pt,escapeinside=||]
def get_tri_area(|\colorbox{orange!30}{points}|):   
    a, b, c = |\colorbox{orange!30}{points}|[0], |\colorbox{orange!30}{points}|[1], |\colorbox{orange!30}{points}|[2]
    |\colorbox{orange!30}{vector1}| = np.array(b) - np.array(a)  
    |\colorbox{orange!30}{vector2}| = np.array(c) - np.array(a)  
    |\colorbox{orange!30}{area\_triangle}| = abs(sp.linalg.norm(sp.cross(vector1,vector2))/2)
    return |\colorbox{orange!30}{area\_triangle}|
        \end{lstlisting}
        \end{minipage}
        \caption{Rewrite variables}
        \label{variable_rewrite}
    \end{subfigure}
    \hspace{0.02\linewidth}
\vrule
\hspace{0.02\linewidth}
\begin{subfigure}[t][0.15\textheight]{0.45\linewidth}
\begin{minipage}[t][0.12\textheight]{\linewidth}
\begin{lstlisting}[language=Python,frame=none,columns=fullflexible,numbers=left,numbersep=5pt,aboveskip=0pt,belowskip=0pt,escapeinside=||]
def get_tri_area(pts):
    a, b, c = pts[0], pts[1], pts[2]
    v1 = [b[i] - a[i] for i in range(len(a))]
    v2 = [c[i] - a[i] for i in range(len(a))]
    area_tri = |\colorbox{orange!30}{math.sqrt}|(pow((v1[0] * v2[1] - v1[1] * v2[0]), 2)) / 2
    return area_tri
\end{lstlisting}
\end{minipage}
\caption{Use different library functions}
\label{library_rewrite}
\end{subfigure}
\hrule
\vspace{0pt}
\begin{subfigure}[t][0.13\textheight]{0.45\linewidth}
\begin{minipage}[t][0.10\textheight]{\linewidth}
\begin{lstlisting}[language=Python,frame=none,columns=fullflexible,numbers=left,numbersep=5pt,aboveskip=0pt,belowskip=0pt,escapeinside=||]
from |\colorbox{orange!30}{scipy.spatial.distance import cdist}|
def get_tri_area(pts):
    a, b, c = pts[0], pts[1], pts[2]
    sides = |\colorbox{orange!30}{cdist}|(pts, pts)
    s= (sides[0][1]+sides[1][2]+sides[2][0]) / 2
    return (s * (s - sides[0][1]) * (s - sides[1][2]) * (s - sides[2][0])) ** 0.5
\end{lstlisting}
\end{minipage}
\caption{Rewrite the code with the same semantics}
\label{semantic_rewrite}
\end{subfigure}
\hspace{0.02\linewidth}
\vrule
\hspace{0.02\linewidth}
\begin{subfigure}[t][0.13\textheight]{0.45\linewidth}
\begin{minipage}[t][0.10\textheight]{\linewidth}
\begin{lstlisting}[language=Python,frame=none,columns=fullflexible,numbers=left,numbersep=5pt,aboveskip=0pt,belowskip=0pt,escapeinside=||]
def get_tri_area(pts):
    v1 = np.array(pts[1]) - np.array(pts[0])
    v2 = np.array(pts[2]) - np.array(pts[0])
    area_tri = abs(sp.linalg.norm(sp.cross(v1, v2)) / 2)
    return area_tri
\end{lstlisting}
\end{minipage}
\caption{Simplify the code}
\label{simplify}
\end{subfigure}
\hrule
\caption{Augmented code samples generated by ChatGPT.}
\label{fig:code_examples}
\end{figure*}

\Fig~\ref{fig:code_examples} shows the five augmented code examples generated by ChatGPT under the guidance of the five proposed code rewriting techniques. In \Fig~\ref{name_rewrite}, it is shown that ChatGPT can understand the meaning of the code and generate augmented data by rewriting the function name from \textit{get\_tri\_area} to \textit{calculate\_triangle\_area}. \Fig~\ref{variable_rewrite} shows that ChatGPT can understand variable abbreviations and convert them into semantically precise variable names, such as converting \textit{pts} to \textit{points}. In \Fig~\ref{library_rewrite}, ChatGPT implements the function using a different library function, \textit{math.sqrt}. 
In \Fig~\ref{semantic_rewrite}, ChatGPT generates enhanced code by using a different library function, \textit{cdst}, and a different mathematical formula while ensuring that the semantics remain unchanged. Finally, in \Fig~\ref{simplify}, we see that ChatGPT simplifies the code snippet by removing statement 2 from the original code.

\subsection{The Data Filtering Stage}
\subsubsection{Bi-encoder \& Cross-encoder}
\begin{figure}[t]
  \centering
  \begin{adjustbox}{max size={0.35\textwidth}{0.35\textheight}}
    \includegraphics[]{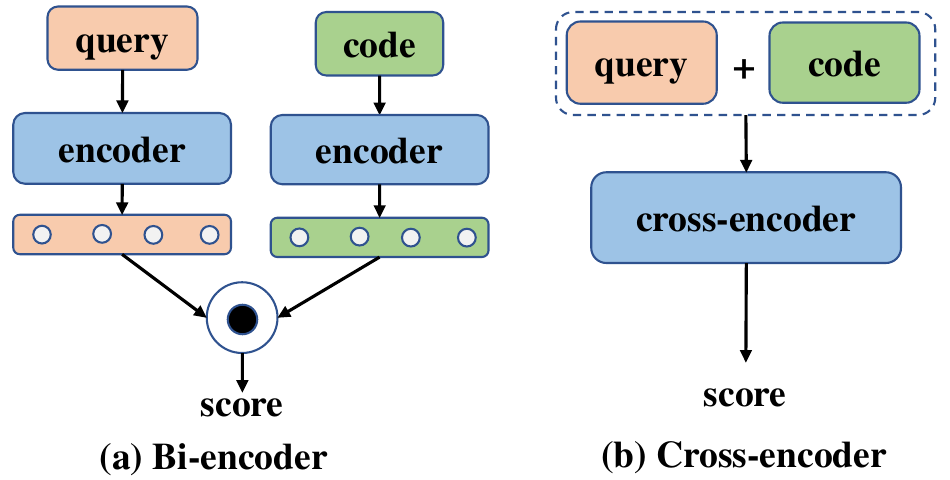}
  \end{adjustbox}
  \caption{The architectures of bi-encoder and cross-encoder.}
  \label{fig:model_architecture}
\end{figure}

In code search, we use deep learning models to score query-code pairs as the basis for their matching. There are two common architectures for code search models: bi-encoder~\cite{gu2018deep,cambronero2019deep,ling2021deep,ye2020leveraging,mamulcs,guo2020graphcodebert,guo2022unixcoder,shi2023cocosoda,wan2019multi,haldar2020multi,gu2021multimodal,shuai2020improving,sun2022code} and cross-encoder~\cite{zhu2020ocor,feng2020codebert,huang2021cosqa,du2021single,ling2020adaptive,shi2022better,li2020learning}. We denote bi-encoder model as $f_{bi}$ and cross-encoder model as $f_{cross}$. As shown in \Fig\ref{fig:model_architecture}(a), given a query-code pair $<Q,C>$, the bi-encoder model encodes the query sequence and code sequence into query vector $\Vec{q}$  and code vector $\Vec{c}$, respectively:\begin{equation}
    \Vec{q} = f_{bi}(Q) \quad \quad \Vec{c} = f_{bi}(C)
\end{equation}And then we calculate the cosine similarity between $\Vec{q}$ and $\Vec{c}$ as the matching score for the code pair:\begin{equation}
    Score_{bi} = sim(\Vec{q},\Vec{c}), \quad sim(\Vec{q},\Vec{c}) = \frac{\Vec{q} \cdot \Vec{c}}{\left \lVert\Vec{q}\right\rVert\left \lVert\Vec{c}\right\rVert}
\end{equation}
For the cross-encoder model shown in \Fig~\ref{fig:model_architecture}(b),  we first concatenate the query sequence $Q$ and code sequence $C$ into a single sequence and then input it into the cross-encoder model to generate the matching score end-to-end:\begin{equation}
    Score_{cross} = f_{cross}([Q,C])
\end{equation}

The bi-encoder model typically has a faster retrieval speed than the cross-encoder model. Assuming there are m query pairs to be searched against a codebase with n code snippets, the bi-encoder model requires m+n model inferences, while the cross-encoder model requires m*n inferences due to the need to concatenate different queries and codes into input sequences. Therefore, the bi-encoder model is commonly used in practical retrieval scenarios. However, the cross-encoder model has better retrieval accuracy than the bi-encoder model. After concatenating the query and code sequences, the cross-encoder model enables token-level interaction between query and code through attention mechanisms, whereas in the bi-encoder model, the interaction between query and code is limited to vector-level. Therefore, the cross-encoder model is considered to be better at matching queries and codes~\cite{hu2023revisiting}. In this paper, we employ the cross-encoder model with better matching capability

as the filtering model to filter the augmented data.

\subsubsection{Filtering Algorithm}After obtaining the augmented data, we filter the data and generate the augmented training dataset using the algorithm shown in Algorithm~\ref{filtering_algorithm}.  
We first train a neural model to measure the semantic relevance between a query and a code snippet, and then filter out code and query pairs with low semantic relevance scores. Specifically,
in line 2, we first train a filtering model $M$ based on cross-encoder architecture on training dataset $D$. In line 3-21, we generate an augmented sample for each query-code pair $<q,c>$ in training dataset $D$. In line 4, we initialize a list to collect the filtered code. In line 5-12, we iterate through the augmented codes and score the original query $q$ and augmented codes using the filtering model. Then we filter out the augmented code $c_{aug}$ with a score below the threshold, and we synthesize augmented samples by combining $q$ and $q_{aug}$. Meanwhile, we add the filtered codes to $code\_list$. Similarly, we filter the queries using the same method in line 13-20. But note that in line 17-18, we generate an augmented sample by combining augmented query $q_{aug}$ and $c_{sample}$ sampled from $code\_list$ randomly to increase the diversity of augmented samples. In our experiments, we have manually checked the quality of the augmented data, and the results show that the augmented code can correctly answer the query. Details can be found in Appendix of replication package\cite{ChatDANCE}.

\label{subsec_data_filter}
\begin{algorithm}

\caption{Filtering Algorithm}
\label{filtering_algorithm}
\renewcommand{\algorithmicrequire}{\textbf{Input:}} 
\renewcommand{\algorithmicensure}{\textbf{Output:}}
\begin{algorithmic}[1]
\REQUIRE original training dataset $D$
\REQUIRE query augmentation dictionary $dict_{q}$ 
\REQUIRE code augmentation dictionary $dict_{c}$,
\REQUIRE filtering threshold for augmented queries $\theta_{q}$ 
\REQUIRE filtering threshold for augmented codes $\theta_{c}$ 
\ENSURE augmented training dataset $D_{aug}$
\STATE Initialize $D_{aug}\leftarrow D$
\STATE Train a filtering model $M$ on dataset $D$
\FOR{$(q,c)$ in $D$}
\STATE $code\_list\leftarrow[c]$
\FOR{$c_{aug}$ in $dict_{c}[c]$}
\STATE $input\_sequence = concatenate(q,c_{aug})$
\STATE $score = M(input\_sequence)$
\IF{$score \ge \theta_{c}$}
\STATE{add $(q,c_{aug})$ to $D_{aug}$}
\STATE{add $c_{aug}$ to $code\_list$}

\ENDIF

\ENDFOR
\FOR{$q_{aug}$ in $dict_{q}[q]$}
\STATE $input\_sequence = concatenate(q_{aug},c)$
\STATE $score = M(input\_sequence)$
\IF{$score \ge \theta_{q}$}
\STATE{$c_{sample} = random.choice(code\_list)$}
\STATE{add $(q_{aug},c_{sample})$ to $D_{aug}$}
\ENDIF

\ENDFOR
\ENDFOR
\RETURN $D_{aug}$
\end{algorithmic}
\end{algorithm}

\subsection{Model Training}
After the data filtering stage, we finetune the bi-encoder model on augmented dataset $D_{aug}$. Following previous studies related to code search~\cite{feng2020codebert,guo2020graphcodebert,guo2022unixcoder,ahmad2021unified,shi2023cocosoda,wang2021codet5}, we finetune model on training dataset by this loss function:\begin{equation}
  L = -\frac{1}{bs}\sum_{i=1}^{bs} \left[\log\frac{(e^{sim(\vec{q_i}, \vec{c_i})/\tau)}}{\sum_{j=1}^{bs} e^{(sim(\vec{q_i}, \vec{c_j})/\tau)}}\right]
\end{equation}where $bs$ denotes the batch size during model training. And $\Vec{q_{i}}$ and $\Vec{c_{i}}$ are vector representations generated from query $q$ and code $c$ using bi-encoder model. The $\tau$ is a hyperparameter. After finetuning the model on the augmented data, we conduct evaluations on the test dataset three times with different random seeds and report the average MRR as the result of the evaluations.

\section{Experimental Design}
\subsection{Evaluated Dataset}
\label{dataset}
We evaluate our approach on a high-quality dataset CoSQA~\cite{huang2021cosqa}. It contains 20,604 web queries collected from the Microsoft Bing search engine and 6,267 Python functions from GitHub. Each instance in CosQA contains a pair of a web query and a code snippet, where one code snippet could be paired with multiple queries. 
Following the original settings in Guo et al.~\cite{guo2022unixcoder}, the training, validation, and testing sets contain 19604, 500, and 500 instances, respectively, and the codebase for code retrieval contains 6,267 code snippets for evaluation. Table~\ref{tab:dataset-table} provides detailed information about the dataset. Additionally, due to the maximum input token length constraint of ChatGPT, we need to ensure that the token lengths of code and query in the CoSQA dataset do not exceed 4096. We conducted a data statistics on the CoSQA dataset and found that the maximum token numbers of codes and queries in the dataset are 1806 and 21, respectively, which do not exceed the token limit (4096) of ChatGPT. More details can be found in Appendix of replication package\cite{ChatDANCE}.


\begin{table}[H]
\centering
\caption{Details of CoSQA dataset.}
\label{tab:dataset-table}
\begin{tabular}{cccc}
\toprule
& \textbf{\# of instances} & \textbf{\# of queries} & \textbf{\# of codes} \\ \midrule
\textbf{Train} & 19604           & 19604         & 6127        \\ 
\textbf{Validation}   & 500             & 500           & 6267        \\ 
\textbf{Test}  & 500             & 500           & 6267        \\ \bottomrule
\end{tabular}
\end{table}

\subsection{Baselines}\label{sec_baseline_method}

We compare \model with two previous data augmentation methods, namely \textbf{QRA} and \textbf{NatGen}, and a strong baseline named Unixcoder~\cite{guo2022unixcoder} in code search to evaluate the effectiveness of \model: 

\begin{enumerate}
\item \textbf{QRA} (Query-Rewritten Augmentation)~\cite{huang2021cosqa} assumes that queries with minor modifications share the same semantics as the original query. Based on this assumption, QRA performs query augmentation by rewriting queries in three ways: deleting a word randomly, copying a word randomly, and switching the position of two words randomly. In our experiments, we use all three ways to perform Query-Rewritten Augmentation.

\item \textbf{NatGen} (De-Naturalizing Source Code)~\cite{chakraborty2022natgen} performs code augmentation by rewriting code based on six semantic-preserving transformations: 
(1) Loop Transformation, (2) DeadCode Injection, (3) Operand Swap, (4) Block Swap, (5) Variable Renaming, (6) Confusing Code Insertion. We use the first five transformations as we find the last transformation is ineffective for the codes in the training set. Similar to code augmentation mentioned in \Sec~\ref{data_aug_via_chatgpt}, we use each transformation rule three times and generate 15 augmented codes in total.
\end{enumerate}

\subsection{Experimental Settings}

During the data augmentation stage, we use the ChatGPT API (GPT-3.5-Turbo-0301)~\cite{chatapi} provided by OpenAI to generate data with default parameter settings. In the subsequent stages, we employed the state-of-the-art model UniXcoder~\cite{guo2022unixcoder} (\Table~\ref{tab:models-table}) as the backbone model in our experiments. It is a Transformer-based architecture with 12 layers, 768 dimensional hidden states, and 12 attention heads.
During the data filtering stage, we take cross-encoder-based Unixcoder as the filtering model and train it using the AdamW optimizer~\cite{loshchilov2017decoupled} with a learning rate of 8e-5 and weight decay of 0.01. We empirically use a filtering score threshold of 0.75 for code and 0.95 for queries during the filtering process. 
During the model training stage, we treat UniXcoder as a bi-encoder model and use AdamW to optimize it with a learning rate of 3e-5 and weight decay of 0.001. %
We conduct experiments three times with different random seeds 
and report the mean values. All experiments are conducted on a machine with 216 GB main memory and Tesla A100 80GB GPU.

\begin{table}[H]
\centering
\caption{Performances of different models on CoSQA.}
\label{tab:models-table}
\begin{NiceTabular}{llc}
\toprule
& \textbf{Model}     & \textbf{MRR}  \\ \midrule
\multirow{3}{*}{\textbf{Encoder-Only}}
&RoBERTa     &60.3 \\ 
&CodeBERT      &65.7 \\ 
&GraphCodeBERT &68.4 \\ \midrule
\multirow{2}{*}{\textbf{Encoder-Decoder}}
& PLBART   & 65.0 \\ 
& CodeT5-base   & 67.8 \\ \midrule
\textbf{Unified} & UniXcoder     & 70.1 \\ \bottomrule
\end{NiceTabular}
\end{table}
\subsection{Evaluation Metrics}
Following previous studies~\cite{gu2018deep,du2021single,shuai2020improving,bui2021self,cambronero2019deep,shi2022better, sun2022code,jain2021contrastive,huang2021cosqa,ling2021deep,wan2019multi,haldar2020multi,li2020learning,ye2020leveraging,mamulcs,zhu2020ocor,gu2021multimodal,ling2020adaptive,wang2021codet5,feng2020codebert,guo2020graphcodebert,guo2022unixcoder,shi2023cocosoda,ahmad2021unified}, we choose two commonly used metrics, namely Mean Reciprocal Rank (MRR) and R@1, to evaluate the performance of the code search models. MRR is the average of reciprocal ranks of the ground truth code snippets for the given queries $Q$. $R@1$ calculates the proportion of queries for which the correct code snippets are ranked first in the returned ranked lists. MRR and R@1 are defined as:

\begin{equation}
MRR = \frac{1}{|Q|} \sum_{i=1}^{|Q|} \frac{1}{rank_{i}}
\end{equation}

\begin{equation}
\label{mrr}
R@1 = \frac{1}{|Q|} \sum_{i=1}^{|Q|} \delta(rank_{i}=1)
\end{equation} 

\noindent where $|Q|$ is the number of queries, $rank_{i}$ is the rank of the correct code for query $i$ and $\delta$ is an indicator function that returns 1 if the input condition is true and 0 otherwise.

\section{Evaluation}

\begin{table}[t]
\caption{Performance of different data augmentation approaches. Standard deviations are shown in parentheses.}  
\centering
\label{tab:method-table}
\begin{tabular}{lll}
\toprule
\textbf{Model} & \textbf{MRR} & \textbf{R@1} \\ \midrule
           UniXcoder & 70.2 (±0.35) & 56.7 (±0.61) \\
   QRA     &   71.3 (±0.84) & 58.6 (±1.04) \\
                    NatGen    & 72.3 (±1.60) & 61.1 (±2.08) \\ \midrule

  \begin{tabular}[c]{@{}c@{}}\textbf{\model}\end{tabular} &
  \begin{tabular}[c]{@{}c@{}}\textbf{75.1 (±0.62)} (↑7.0\%)\end{tabular} &
  \begin{tabular}[c]{@{}c@{}}\textbf{64.2 (±0.87)} (↑13.2\%)\end{tabular} \\ \bottomrule
\end{tabular}
\end{table}

\subsection{RQ1: What Is the Effectiveness of Our Approach?}
\label{sec_effectiveness}

\textbf{Overall results. }

We evaluate the effectiveness of \model by comparing it with baselines on the CoSQA~\cite{huang2021cosqa} dataset with two common metrics MRR and R@1. The experiment results are shown in \Table~\ref{tab:method-table}. From \Table~\ref{tab:method-table}, we can see that our model outperforms all baselines. Specifically, our approach improved the base model (UniXcoder) by 7.0\% in MRR and 13.2\% in R@1. Furthermore, our approach has smaller standard deviations in both MRR and R@1 metrics compared to the baselines, indicating better stability. Overall, our approach performs the best among all models

\textbf{Case study. } 

\Fig~\ref{fig:case2} presents the top-1 code snippets returned by QRA, NatGen, and \model for the query ``how to remove blank lines from a text file in python''. 
We can see that~\model returns the correct code snippet, which reads the file content and removes blank lines. In contrast, the code snippet returned by QRA and NatGen is incorrect, which can remove blank lines but fails to operate on a file. 

\begin{figure}[H]
    \centering
    \begin{subfigure}{0.45\textwidth}
        \begin{lstlisting}[caption={The top-1 code returned by NatGen and QRA}, label=lst:natgen]
def lines(input):
    for raw_line in input:
        line = raw_line.strip()
        if line and not line.startswith('#'):
            yield strip_comments(line)
        \end{lstlisting}
        \label{fig:to_int64}
    \end{subfigure}
    \hfill
    \begin{subfigure}{0.45\textwidth}
        \begin{lstlisting}[caption={The top-1 code returned by \model}, label=lst:scale_dtype]
def get_stripped_file_lines(filename):
    try:
        lines = open(filename).readlines()
    except FileNotFoundError:
        fatal("Could not open file: {!r}".format(filename))
    return [line.strip() for line in lines]
        \end{lstlisting}
        \label{fig:scale_dtype}
    \end{subfigure}
    \caption{The top-1 code returned by QRA, NatGen, and \model for the query ``how to remove blank lines from a text file in python''.}
    \label{fig:case2}
\end{figure}

\begin{center}
    \begin{tcolorbox}[colback=gray!10!white,colframe=black,coltext=black,boxsep=-3pt]
    \textbf{Summary:} 
    \model significantly outperforms baselines on code search and the case study intuitively demonstrates the superiority of \model.
    \end{tcolorbox}
\end{center}

\subsection{RQ2: How Much Do Different Components Contribute?}
\label{sec_abalation}

As described in \Sec\ref{approach_sec}, \model contains 3 main components: query augmentation, code augmentation, and data filtering. We conduct an ablation study by removing each component at a time.When examining the impact of data filtering component, we explore three experimental settings: (1) no filtering on queries, (2) no filtering on code, and (3) no filtering on both. The results are shown in \Table~\ref{tab:ablation-table}.

From \Table~\ref{tab:ablation-table}, we can see that the performance of \model decreases when any individual component is removed. This indicates that each component of \model plays an important role in overall performance improvement.
Furthermore, we observe that data filtering is critical to the model's performance and stability. Without data filtering, they would be significantly affected. The filtering mechanism can effectively eliminate noise within the data, resulting in high-quality data that is essential for improving model performance.

\begin{table}[t]
\centering
\caption{Ablation study results of \model.}
\label{tab:ablation-table}
\begin{tabular}{lcc}
\toprule
\textbf{Model}          & \textbf{MRR}         & \textbf{R@1}         \\ \midrule
\textbf{\model}          & \textbf{75.1 (±0.62)}  & \textbf{64.2 (±0.87)} \\
\ \ w/o query aug            & 74.6 (±0.66)          & 63.3 (±1.41)          \\
\ \ w/o code aug             & 72.2 (±0.65)          & 59.5 (±2.00)          \\
\ \ w/o query filtering      & 73.4 (±0.40)          & 61.9 (±0.50)          \\
\ \ w/o code filtering       & 73.5 (±0.30)          & 61.5 (±0.75)          \\
\ \ w/o any filtering        & 72.9 (±2.15)          & 61.3 (±4.02)          \\ \bottomrule
\end{tabular}
\end{table}

\begin{center}
    \begin{tcolorbox}[colback=gray!10!white,colframe=black,coltext=black,boxsep=-3pt]
    \textbf{Summary:} The ablation experiments demonstrate the effectiveness of each component of \model.
    \end{tcolorbox}
\end{center}

\subsection{RQ3: What Is the Impact of Different Hyperparameters?}
\label{sec_hyperparameters}

\begin{figure}[t]
  \centering
  \captionsetup[subfigure]{font=large}
  \begin{adjustbox}{max width=0.5\textwidth,center}
    \begin{subfigure}{0.5\textwidth}
      \centering
      \includegraphics[trim={0.5cm 0.5cm 1cm 1cm},clip,width=\linewidth]{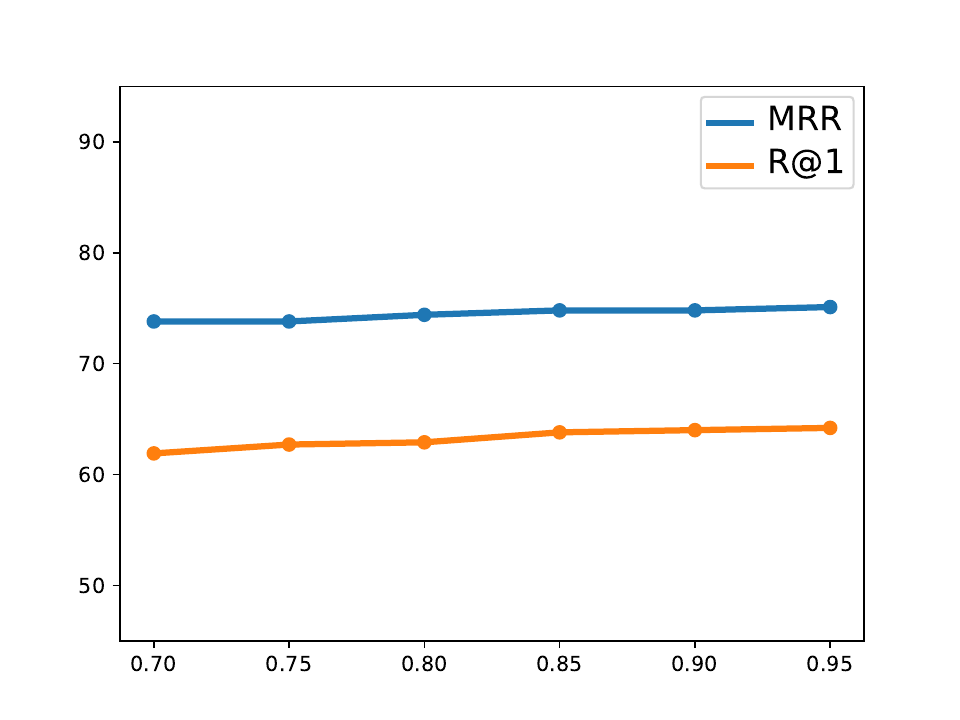}
      \caption{Threshold for filtering query $\theta_{q}$}
      \label{fig:theta_q}
    \end{subfigure}%
    \hspace{0em}%
    \begin{subfigure}{0.5\textwidth}
      \centering
      \includegraphics[trim={0.5cm 0.5cm 1cm 1cm},clip,width=\linewidth]{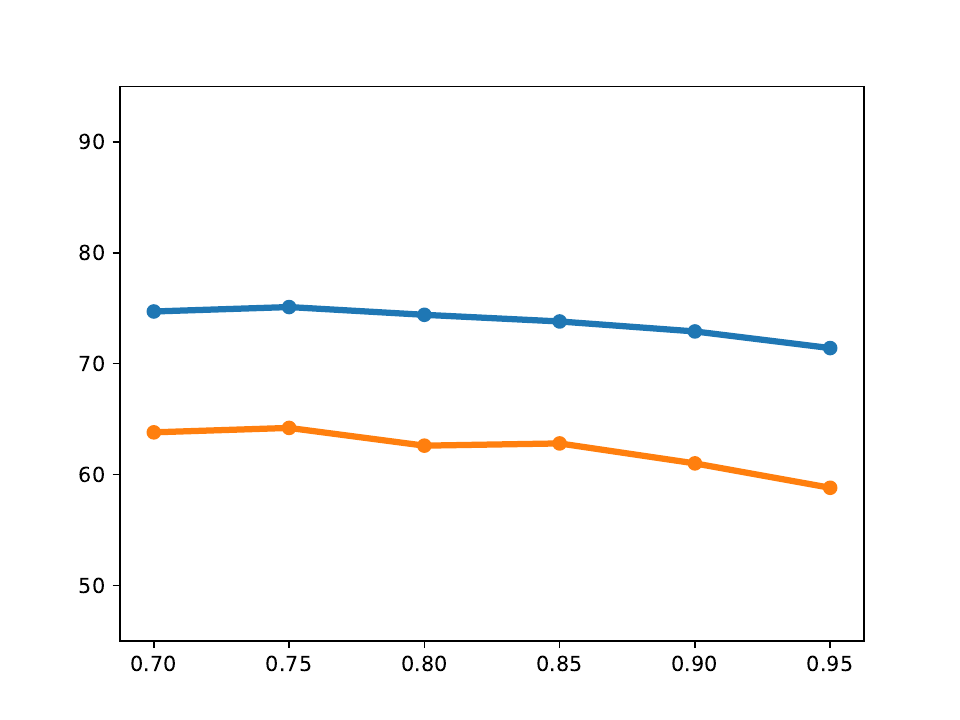}
      \caption{Threshold for filtering code $\theta_{c}$}
      \label{fig:theta_c}
    \end{subfigure}
  \end{adjustbox}
  
  \begin{adjustbox}{max width=0.5\textwidth,center}
    \begin{subfigure}{0.5\textwidth}
      \centering
      \includegraphics[trim={0.5cm 0.5cm 1cm 1cm},clip,width=\linewidth]{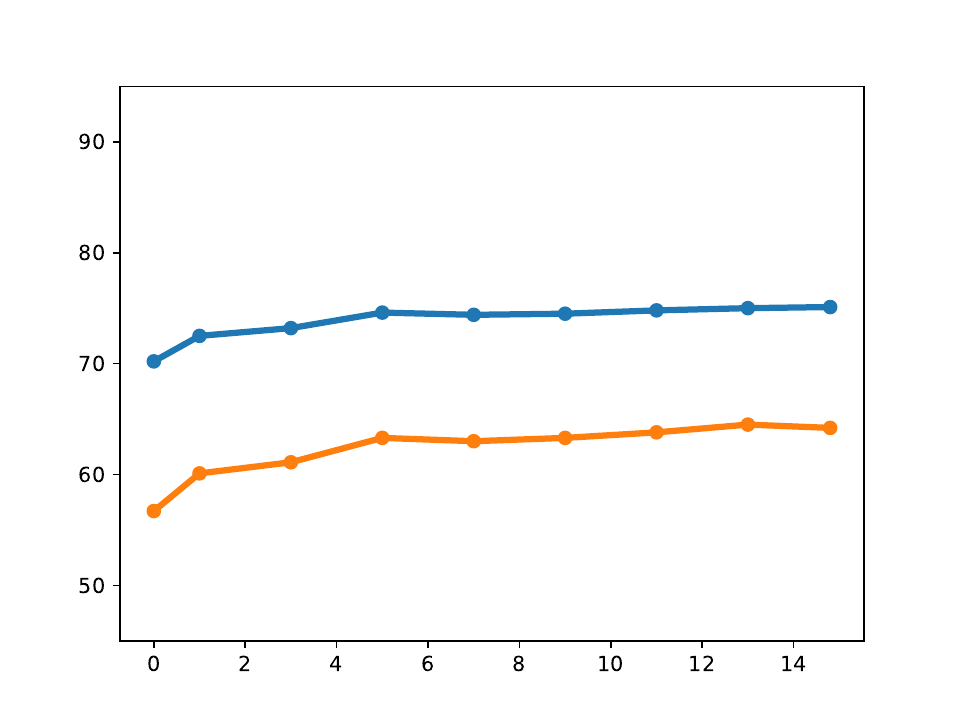}
      \caption{{Average number of augmented sample $N_{aug}$}}
      \label{fig:number}
    \end{subfigure}%
    \hspace{0em}%
    \begin{subfigure}{0.5\textwidth}
      \centering
      \includegraphics[trim={0.5cm 0.5cm 1cm 1cm},clip,width=\linewidth]{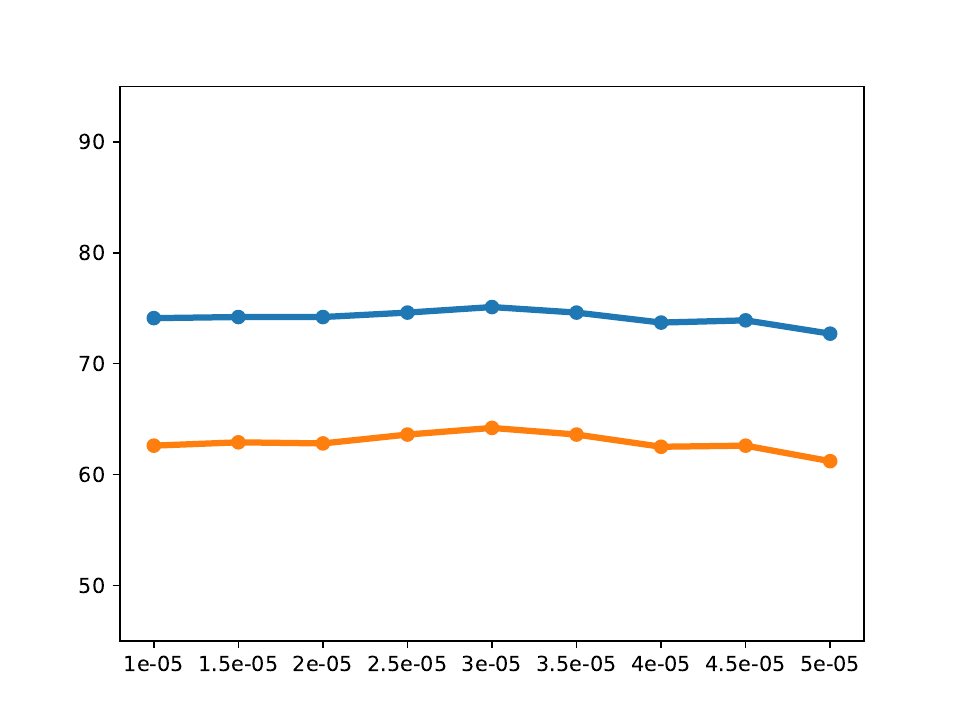}
      \caption{Learning rate {$lr$}}
      \label{fig:lr}
    \end{subfigure}
  \end{adjustbox}
  
  \caption{The impact of different hyperparameters.}
  \label{fig:impact}
\end{figure}

We investigate the impact of hyper-parameters including query filtering threshold $\theta_{q}$, code filtering threshold $\theta_{c}$, the average number of augmentations per sample $N_{aug}$, and learning rate $lr$. We conduct the experiments within ranges surrounding the default values, and the results are shown in \Fig~\ref{fig:impact}.The results show that the performance preserves stable as $\theta_{q}$ varies from 0.7 to 0.95. Similarly,
the results show that the performance remains stable when $\theta_c$ varies from 0.7 to 0.9. However, we observe a sharp decline in performance when $\theta_c$ changes from 0.9 to 0.95. This is because when $\theta_c$ is set to 0.95, a large number of code augmentation samples are filtered out, leading to a significant decrease in the total number of samples and thus a decline in performance. In \Fig~\ref{fig:number}, we find that the performance improves significantly as the average number of augmented samples per original sample increases from 0 to 5. Beyond this point, the performance improvement gradually stabilizes. Finally, from \Fig~\ref{fig:lr}, we observe that \model is stable when learning rate varies from 1e-5 to 5e-5.

\begin{center}
    \begin{tcolorbox}[colback=gray!10!white,colframe=black,coltext=black,boxsep=-3pt]
    \textbf{Summary:} %
    Overall, \model performs stably across a range of  hyper-parameter values ($0.7 \leq \theta_c \leq 0.9$, $0.7 \leq \theta_q \leq 0.95$, $N_{aug} \geq 5$, $1e^{-5} \leq lr  \leq 5e^{-5}$).
    \end{tcolorbox}
\end{center}

\subsection{RQ4: Why Does Our Approach Work?}
\label{sec_analysis}

In general, the primary advantage of our approach is generating a substantial volume of high-quality training data applicable to both queries and code snippets, which enables the model to achieve superior performance. By augmenting queries, we can introduce richer syntactic structures and more expressive forms, enhancing the model's robustness and generalization ability. Additionally, by augmenting codes, we can enable the model to learn more complex syntactic structures and semantic information in code snippets, improving the model's understanding of code snippets. This method not only amplifies the training data quantity but also elevates its quality, leading to a more effective and robust model. 
Next, 
we conduct quantitative and qualitative analysis to investigate why \model works well in detail. 

\subsubsection{Quantitative Analysis}
We investigate why \model works by studying the distribution of data representations learned by models. We use $\ell_{align}$ and $\ell_{uniformity}$~\cite{wang2020understanding} metrics to evaluate the quality of the representations learned by models, which are widely used in contrastive learning~\cite{wang2020understanding,gao2021simcse, Meng2021coco,wang2021understanding}. The $\ell_{align}$ and $\ell_{uniformity}$ metrics are defined as follows in \Eq~\ref{loss_uniform_align}:
\begin{equation}
\label{loss_uniform_align}
\begin{aligned}
&\ell_{align} = \underset{(x,y) \sim D_{pair}}{\mathbb{E}} [\lVert f(x)-f(y) \rVert^\alpha_{2}], \quad \alpha \textgreater{} 0 \\
&\ell_{uniformity} = log \underset{(x,y) \sim D }{\mathbb{E}} e^{-t\lVert f(x)-f(y) \rVert^{2}_{2}}, \quad t \textgreater{} 0
\end{aligned}
\end{equation}
where $(x,y) \sim D_{pair}$ means that $x$ and $y$ (such as query and code) are paired, and $(x,y) \sim D$ means that $x$ and $y$ are independently and identically distributed. $f(x)$ and $f(y)$ represent the representations learned by the model, and $\lVert f(x)-f(y) \rVert_{2}$ represents the 2-norm distance between the representations. The hyperparameters $\alpha$ and $t$ are set to 2 in our experiments. The $\ell_{alignment}$ is defined as the expected distance between paired representations, which measures the degree of matching between paired representations. From \Eq~\ref{loss_uniform_align}, we know that the smaller the distance between paired representations, the closer the alignment loss is to 0. In the extreme case, if the distance between all paired representations is 0, the alignment loss is 0. The $\ell_{uniformity}$ measures the uniformity of the distribution of representations. From \Eq~\ref{loss_uniform_align}, we know that the closer the distribution of representations is to uniformity, the closer the value of uniformity loss is to negative infinity. According to~\cite{wang2020understanding,gao2021simcse}, better alignment and uniformity can enable the model to have better performance and generalization. In our experiments, we use $\ell_{alignment}$ and $\ell_{uniformity}$ to measure these two properties. The lower the values of two losses, the better the alignment and uniformity of the learned representations.

\begin{figure}[t]
  \centering
  \includegraphics[width=0.5\textwidth]{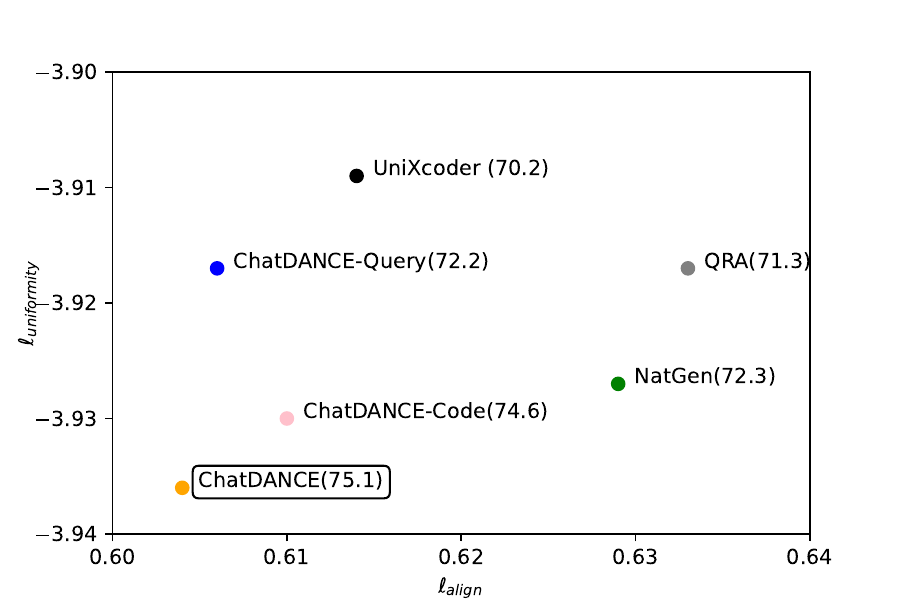}
   \caption{$\ell_{align}$-$\ell_{uniformity}$ plot of different models. \model-Query and \model-Code  represent performing data augmentation on queries or codes, respectively. %
   }
  \label{fig:loss}
\end{figure}

The $\ell_{align}$-$\ell_{uniformity}$ plot shown in \Fig~\ref{fig:loss} reveals several observations. Firstly, our approach reduces both $\ell_{alignment}$ and $\ell_{uniformity}$ compared to the two baselines. This implies that our approach can enhance both the alignment and uniformity of the learned representations, resulting in superior performance and generalization. In contrast, the baselines only reduce the uniformity loss but increase the alignment loss, indicating a deterioration in the alignment of the learned representations. Therefore, we believe that better alignment and uniformity are crucial factors in achieving superior performance and generalization in our approach. Secondly, our approach can simultaneously improve the alignment and uniformity of the learned representations, even when data augmentation is applied only to either code or query. In contrast, the baselines fail to achieve this. We attribute this success to our approach's ability to generate a large amount of high-quality and diverse data, thereby improving uniformity while ensuring the quality and diversity of the generated data. This allows the model to better comprehend the semantic meaning of both query and code, ultimately improving the alignment of the learned representations.
    
\subsubsection{Qualitative Analysis}
\begin{figure}[t]
  \centering
  \captionsetup[subfigure]{font=large}
  \begin{adjustbox}{max width=0.5\textwidth,center}
    \begin{subfigure}{0.5\textwidth}
      \centering
      \includegraphics[trim={0.5cm 0.5cm 1cm 1cm},clip,width=\linewidth]{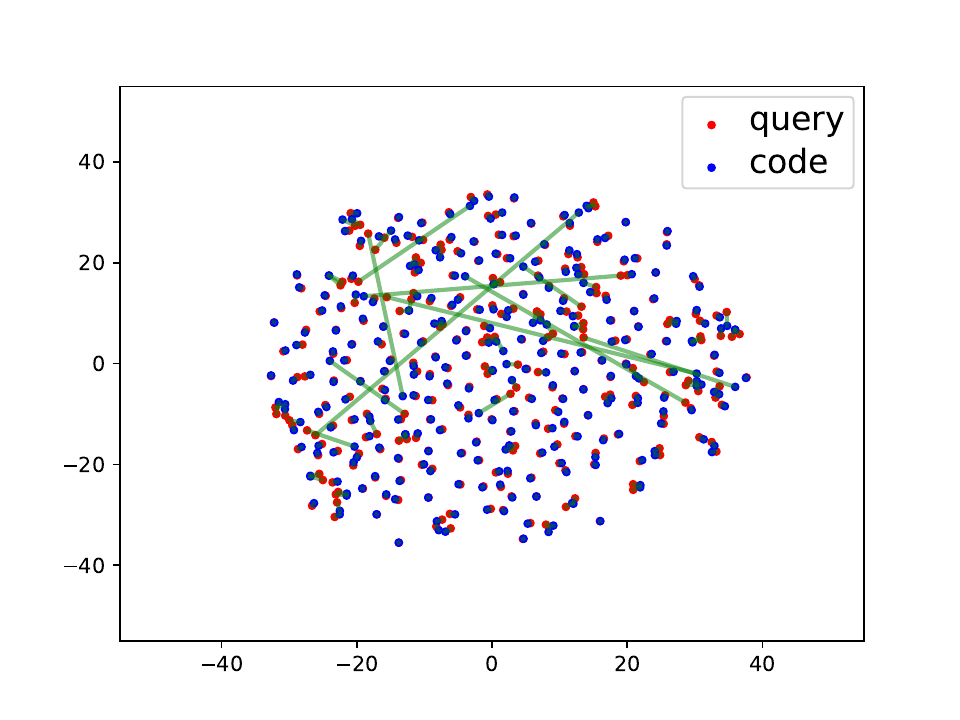}
      \caption{UniXcoder}
      \label{fig:unixcoder}
    \end{subfigure}%
    \hspace{0em}%
    \begin{subfigure}{0.5\textwidth}
      \centering
      \includegraphics[trim={0.5cm 0.5cm 1cm 1cm},clip,width=\linewidth]{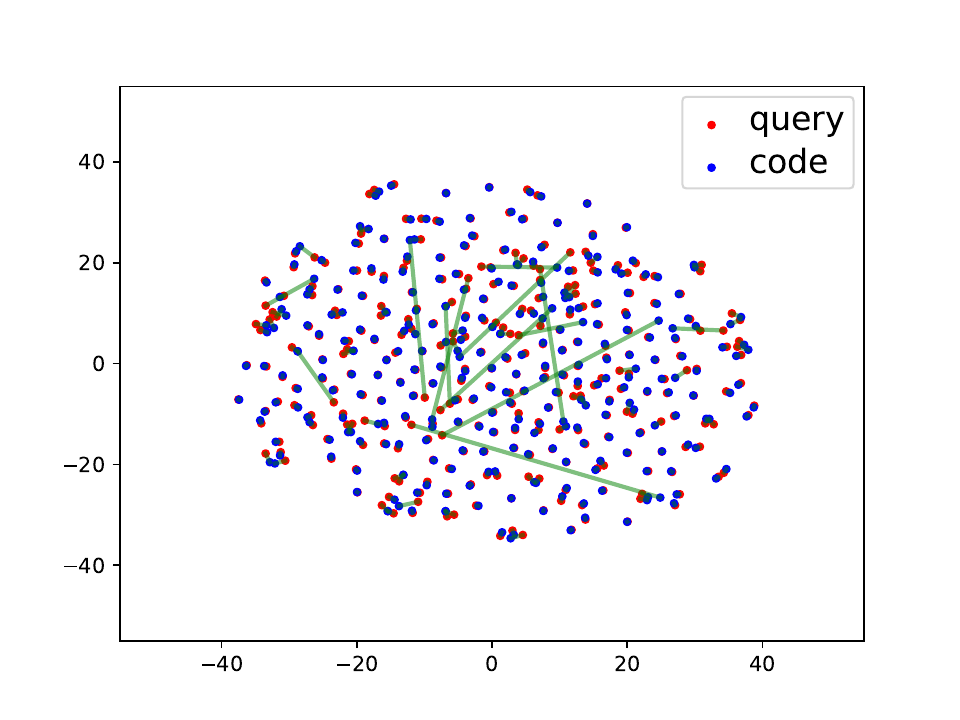}
      \caption{QRA}
      \label{fig:qra}
    \end{subfigure}
  \end{adjustbox}
  \begin{adjustbox}{max width=0.5\textwidth,center}
    \begin{subfigure}{0.5\textwidth}
      \centering
      \includegraphics[trim={0.5cm 0.5cm 1cm 1cm},clip,width=\linewidth]{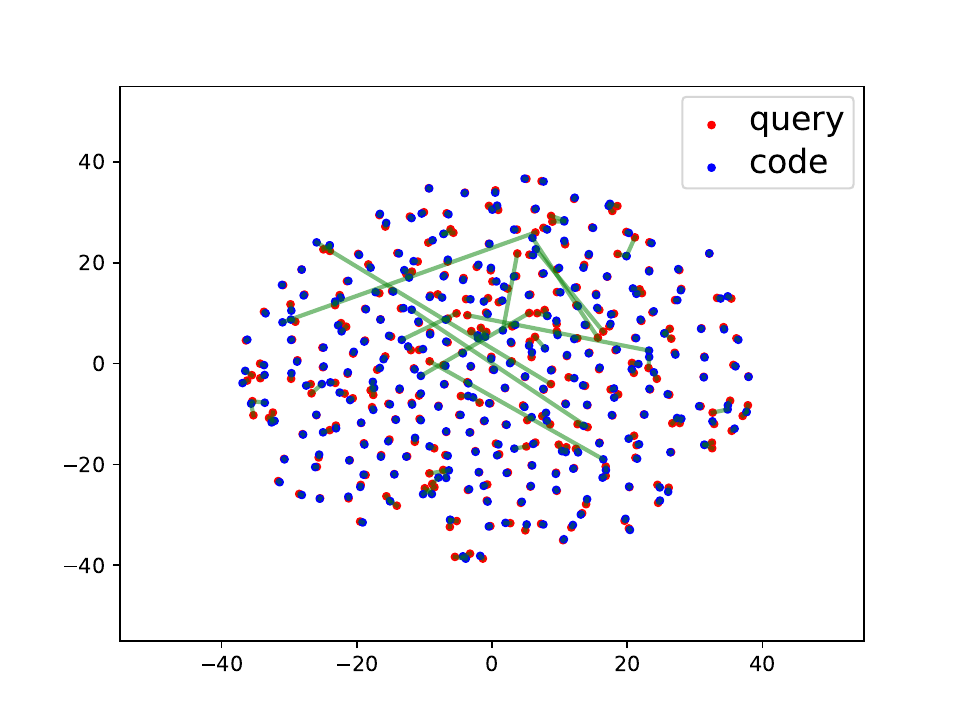}
      \caption{NatGen}
      \label{fig:natgen}
    \end{subfigure}%
    \hspace{0em}%
    \begin{subfigure}{0.5\textwidth}
      \centering
      \includegraphics[trim={0.5cm 0.5cm 1cm 1cm},clip,width=\linewidth]{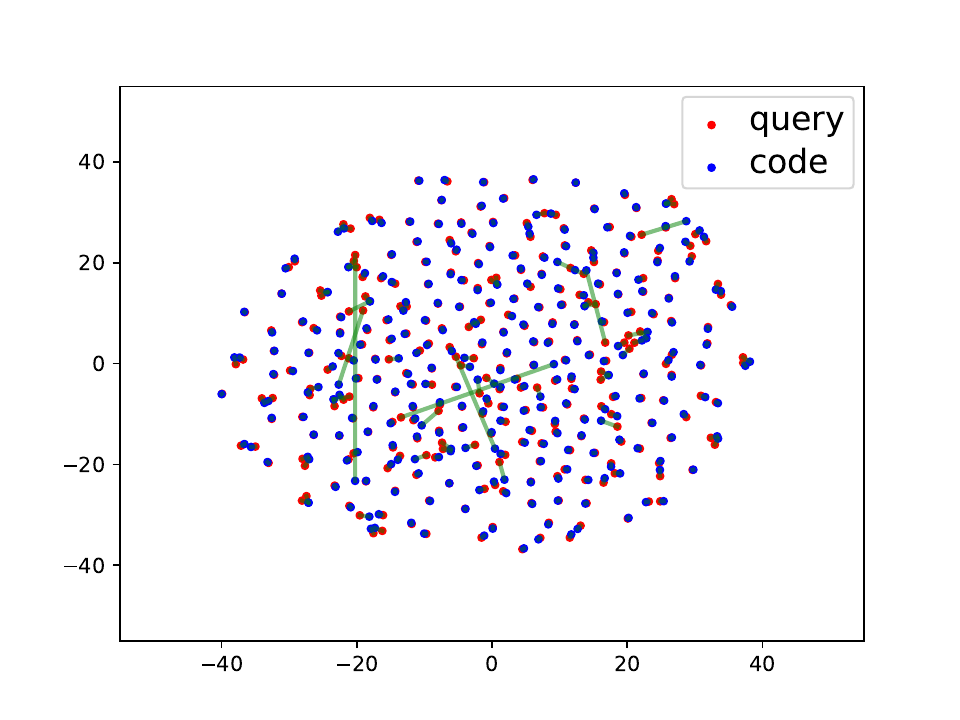}
      \caption{\model}
      \label{fig:chatgpt}
    \end{subfigure}
  \end{adjustbox}
  
  \caption{t-SNE visualization of representations of queries and code snippets. 
  The red and blue dots represent queries and code, respectively. The green line connecting red and blue dot shows the distance between query-code pairs.}
  \label{fig:comparison}
\end{figure}

We visualize the distribution of representations learned by four models shown in \Fig~\ref{fig:comparison} to intuitively explore why our approach works. First, we sample 300 query-code pairs from the test set of CosQA and obtain their representations in the high-dimensional vector space by feeding them into the model. Then, we use t-SNE~\cite{van2008visualizing} to perform dimensionality reduction on the representations and visualize their distribution. The experimental results are shown in \Fig~\ref{fig:comparison}. We visualized the distribution of representations learned by four models: UniXcoder, QRA, NatGen, and \model, which are shown in \Fig~\ref{fig:unixcoder}, \ref{fig:qra}, \ref{fig:natgen}, and \ref{fig:chatgpt}, respectively. The red dots represent the code, and the blue dots represent the query. The lines between the red and blue dots indicate the distance between the code and the query.

From the experimental results, we can observe that: (1) In \Fig~\ref{fig:unixcoder}, \ref{fig:qra}, \ref{fig:natgen}, and \ref{fig:chatgpt}, most of the green lines are very short, indicating that for most query-code pairs, the model is able to align their representations in the high-dimensional vector space to close locations. (2) In \Fig~\ref{fig:comparison}, compared to the other three figures, \Fig~\ref{fig:chatgpt} has the fewest long green lines. This suggests that our approach can enable the model to have better alignment compared to the other methods, as there is the fewest number of representations with long distances. Overall, the visualization results intuitively demonstrate that our approach can enable the model to learn better representations compared to the baselines by improving the alignment and uniformity of the learned representations.

\begin{center}
    \begin{tcolorbox}[colback=gray!10!white,colframe=black,coltext=black,boxsep=-3pt]
    \textbf{Summary:} \model learns a more uniform distribution of representations and effectively aligns the learned representations of paired code snippets and queries.
    \end{tcolorbox}
\end{center}

\section{Discussion}
\subsection{Discussion on Using LLMs}

Generally, LLMs such as ChatGPT\cite{chatgpt} are commonly used for generation tasks. However, when used for code search, a smaller code search model is more efficient. In our pre-study, we find that ChatGPT takes an average of 5 seconds to generate a single code snippet, while a code search model retrieves code with an average time of 0.023 seconds. As a result, code search models demonstrate higher efficiency compared to directly using LLMs. Additionally, when the codebase is evolving, training ChatGPT to update its knowledge is costly, whereas training dedicated code search models on the updated codebase is relatively inexpensive. In general, using smaller models for code search is more efficient and cost-effective compared to using LLMs directly.

\subsection{Metric Choice}

In our experiments, we choose MRR and R@1 as the evaluation metrics instead of R@5 and R@10, even though MRR and R@k (where k=1, 5, 10) are commonly used metrics in code search. This is because our experiments employed a powerful baseline, UniXcoder, which achieved MRR scores greater than 0.6. This suggests that for the majority of queries, the correct code is ranked within the top-3 results. Consequently, R@5 and R@10 metrics are not very informative in this case to reflect performance improvements. Therefore, we focus on using MRR and R@1 as the primary evaluation metrics. However, for completeness and comparison purposes, we also provide the evaluation results under R@5 and R@10 in Appendix of replication package \cite{ChatDANCE}.

\subsection{Threats To Validity}
We identify the following threats to our approach:

\textbf{LLM Choice.} In our experiments, we only use the GPT-3.5-Turbo-0301 model~\cite{chatapi} to perform data augmentation. This is because OpenAI only made this API available during the experiment. Additionally, other large open-source models such as Llama~\cite{touvron2023llama} were not open-sourced at the time, so we did not explore the effectiveness of our method using more LLMs. In the future, we will combine our method with more LLMs to comprehensively explore its effectiveness.

\textbf{Dataset Choice.} As shown in \Sec~\ref{dataset}, we use the CoSQA~\cite{huang2021cosqa} dataset instead of other datasets such as  CodeSearchNet~\cite{husain2019codesearchnet} in our experiments. This is because CoSQA is collected from real-world queries and with manually checked data quality, which is advantageous for us to explore the effectiveness of our method on both query and code enhancement. Additionally, CodeSearchNet has a much larger dataset size compared to CoSQA, and considering that our method is time-consuming, performing data augmentation on CodeSearchNet would require a significant amount of time. 
In the future, we will explore the effectiveness of our method on more datasets.

\section{Conclusion}
In this paper, we explore the effectiveness of ChatGPT-based data augmentation in the code search task and demonstrate its effectiveness through extensive experiments. We propose a novel and efficient data augmentation method called \model, which generates a large amount of high-quality data by rewriting both code and queries using ChatGPT with carefully designed guidance. Additionally, we introduced a filtering mechanism that removes low-quality data from the augmented data, further enhancing the quality of the augmented data. Our experimental results demonstrate that our method can effectively improve the performance of code search models and significantly outperform the baselines. We believe that our augmentation method could be adapted to other code intelligence tasks such as code summarization and different programming languages. Furthermore, our exploration on prompt engineering for code search may inspire researchers to effectively leverage LLMs in solving various software engineering tasks. Replication package is anonymously available at \url{https://anonymous.4open.science/r/ChatDANCE}.

\section*{Acknowledgement}
The work described in this paper was supported by the National Natural Science Foundation of China (No.62002352) and the National Natural Science  Foundation of China (No.62032025).

\bibliography{reference}

\begin{thebibliography}{10}
\providecommand{\url}[1]{#1}
\csname url@samestyle\endcsname
\providecommand{\newblock}{\relax}
\providecommand{\bibinfo}[2]{#2}
\providecommand{\BIBentrySTDinterwordspacing}{\spaceskip=0pt\relax}
\providecommand{\BIBentryALTinterwordstretchfactor}{4}
\providecommand{\BIBentryALTinterwordspacing}{\spaceskip=\fontdimen2\font plus
\BIBentryALTinterwordstretchfactor\fontdimen3\font minus \fontdimen4\font\relax}
\providecommand{\BIBforeignlanguage}[2]{{%
\expandafter\ifx\csname l@#1\endcsname\relax
\typeout{** WARNING: IEEEtran.bst: No hyphenation pattern has been}%
\typeout{** loaded for the language `#1'. Using the pattern for}%
\typeout{** the default language instead.}%
\else
\language=\csname l@#1\endcsname
\fi
#2}}
\providecommand{\BIBdecl}{\relax}
\BIBdecl

\bibitem{github}
``Github,'' \url{https://github.com/}.

\bibitem{allamanis2018survey}
M.~Allamanis, E.~T. Barr, P.~Devanbu, and C.~Sutton, ``A survey of machine learning for big code and naturalness,'' \emph{ACM Computing Surveys (CSUR)}, vol.~51, no.~4, pp. 1--37, 2018.

\bibitem{singer2010examination}
J.~Singer, T.~Lethbridge, N.~Vinson, and N.~Anquetil, ``An examination of software engineering work practices,'' in \emph{CASCON First Decade High Impact Papers}, 2010, pp. 174--188.

\bibitem{nie2016query}
L.~Nie, H.~Jiang, Z.~Ren, Z.~Sun, and X.~Li, ``Query expansion based on crowd knowledge for code search,'' \emph{IEEE Transactions on Services Computing}, vol.~9, no.~5, pp. 771--783, 2016.

\bibitem{mcmillan2011portfolio}
C.~McMillan, M.~Grechanik, D.~Poshyvanyk, Q.~Xie, and C.~Fu, ``Portfolio: finding relevant functions and their usage,'' in \emph{Proceedings of the 33rd International Conference on Software Engineering}, 2011, pp. 111--120.

\bibitem{lu2015query}
M.~Lu, X.~Sun, S.~Wang, D.~Lo, and Y.~Duan, ``Query expansion via wordnet for effective code search,'' in \emph{2015 IEEE 22nd International Conference on Software Analysis, Evolution, and Reengineering (SANER)}.\hskip 1em plus 0.5em minus 0.4em\relax IEEE, 2015, pp. 545--549.

\bibitem{lv2015codehow}
F.~Lv, H.~Zhang, J.-g. Lou, S.~Wang, D.~Zhang, and J.~Zhao, ``Codehow: Effective code search based on api understanding and extended boolean model (e),'' in \emph{2015 30th IEEE/ACM International Conference on Automated Software Engineering (ASE)}.\hskip 1em plus 0.5em minus 0.4em\relax IEEE, 2015, pp. 260--270.

\bibitem{linstead2009sourcerer}
E.~Linstead, S.~Bajracharya, T.~Ngo, P.~Rigor, C.~Lopes, and P.~Baldi, ``Sourcerer: mining and searching internet-scale software repositories,'' \emph{Data Mining and Knowledge Discovery}, vol.~18, pp. 300--336, 2009.

\bibitem{gu2018deep}
X.~Gu, H.~Zhang, and S.~Kim, ``Deep code search,'' in \emph{Proceedings of the 40th International Conference on Software Engineering}, 2018, pp. 933--944.

\bibitem{cambronero2019deep}
J.~Cambronero, H.~Li, S.~Kim, K.~Sen, and S.~Chandra, ``When deep learning met code search,'' in \emph{Proceedings of the 2019 27th ACM Joint Meeting on European Software Engineering Conference and Symposium on the Foundations of Software Engineering}, 2019, pp. 964--974.

\bibitem{ling2021deep}
X.~Ling, L.~Wu, S.~Wang, G.~Pan, T.~Ma, F.~Xu, A.~X. Liu, C.~Wu, and S.~Ji, ``Deep graph matching and searching for semantic code retrieval,'' \emph{ACM Transactions on Knowledge Discovery from Data (TKDD)}, vol.~15, no.~5, pp. 1--21, 2021.

\bibitem{shuai2020improving}
J.~Shuai, L.~Xu, C.~Liu, M.~Yan, X.~Xia, and Y.~Lei, ``Improving code search with co-attentive representation learning,'' in \emph{Proceedings of the 28th International Conference on Program Comprehension}, 2020, pp. 196--207.

\bibitem{du2021single}
L.~Du, X.~Shi, Y.~Wang, E.~Shi, S.~Han, and D.~Zhang, ``Is a single model enough? mucos: A multi-model ensemble learning approach for semantic code search,'' in \emph{Proceedings of the 30th ACM International Conference on Information \& Knowledge Management}, 2021, pp. 2994--2998.

\bibitem{li2020learning}
W.~Li, H.~Qin, S.~Yan, B.~Shen, and Y.~Chen, ``Learning code-query interaction for enhancing code searches,'' in \emph{2020 IEEE International Conference on Software Maintenance and Evolution (ICSME)}.\hskip 1em plus 0.5em minus 0.4em\relax IEEE, 2020, pp. 115--126.

\bibitem{ye2020leveraging}
W.~Ye, R.~Xie, J.~Zhang, T.~Hu, X.~Wang, and S.~Zhang, ``Leveraging code generation to improve code retrieval and summarization via dual learning,'' in \emph{Proceedings of The Web Conference 2020}, 2020, pp. 2309--2319.

\bibitem{mamulcs}
Y.~Ma, Y.~Yu, S.~Li, Z.~Jia, J.~Ma, R.~Xu, W.~Dong, and X.~Liao, ``Mulcs: Towards a unified deep representation for multilingual code search.''

\bibitem{zhu2020ocor}
Q.~Zhu, Z.~Sun, X.~Liang, Y.~Xiong, and L.~Zhang, ``Ocor: an overlapping-aware code retriever,'' in \emph{Proceedings of the 35th IEEE/ACM International Conference on Automated Software Engineering}, 2020, pp. 883--894.

\bibitem{wan2019multi}
Y.~Wan, J.~Shu, Y.~Sui, G.~Xu, Z.~Zhao, J.~Wu, and P.~Yu, ``Multi-modal attention network learning for semantic source code retrieval,'' in \emph{2019 34th IEEE/ACM International Conference on Automated Software Engineering (ASE)}.\hskip 1em plus 0.5em minus 0.4em\relax IEEE, 2019, pp. 13--25.

\bibitem{haldar2020multi}
R.~Haldar, L.~Wu, J.~Xiong, and J.~Hockenmaier, ``A multi-perspective architecture for semantic code search,'' in \emph{Annual Meeting of the Association for Computational Linguistics}, 2020.

\bibitem{gu2021multimodal}
J.~Gu, Z.~Chen, and M.~Monperrus, ``Multimodal representation for neural code search,'' in \emph{2021 IEEE International Conference on Software Maintenance and Evolution (ICSME)}.\hskip 1em plus 0.5em minus 0.4em\relax IEEE, 2021, pp. 483--494.

\bibitem{ling2020adaptive}
C.~Ling, Z.~Lin, Y.~Zou, and B.~Xie, ``Adaptive deep code search,'' in \emph{Proceedings of the 28th International Conference on Program Comprehension}, 2020, pp. 48--59.

\bibitem{sun2022code}
W.~Sun, C.~Fang, Y.~Chen, G.~Tao, T.~Han, and Q.~Zhang, ``Code search based on context-aware code translation,'' in \emph{Proceedings of the 44th International Conference on Software Engineering}, 2022, pp. 388--400.

\bibitem{shi2022better}
Y.~Shi, Y.~Yin, Z.~Wang, D.~Lo, T.~Zhang, X.~Xia, Y.~Zhao, and B.~Xu, ``How to better utilize code graphs in semantic code search?'' in \emph{Proceedings of the 30th ACM Joint European Software Engineering Conference and Symposium on the Foundations of Software Engineering}, 2022, pp. 722--733.

\bibitem{wang2021codet5}
Y.~Wang, W.~Wang, S.~Joty, and S.~C. Hoi, ``Codet5: Identifier-aware unified pre-trained encoder-decoder models for code understanding and generation,'' in \emph{Proceedings of the 2021 Conference on Empirical Methods in Natural Language Processing}, 2021, pp. 8696--8708.

\bibitem{feng2020codebert}
Z.~Feng, D.~Guo, D.~Tang, N.~Duan, X.~Feng, M.~Gong, L.~Shou, B.~Qin, T.~Liu, D.~Jiang \emph{et~al.}, ``Codebert: A pre-trained model for programming and natural languages,'' in \emph{Findings of the Association for Computational Linguistics: EMNLP 2020}, 2020, pp. 1536--1547.

\bibitem{guo2020graphcodebert}
D.~Guo, S.~Ren, S.~Lu, Z.~Feng, D.~Tang, L.~Shujie, L.~Zhou, N.~Duan, A.~Svyatkovskiy, S.~Fu \emph{et~al.}, ``Graphcodebert: Pre-training code representations with data flow,'' in \emph{International Conference on Learning Representations}.

\bibitem{guo2022unixcoder}
D.~Guo, S.~Lu, N.~Duan, Y.~Wang, M.~Zhou, and J.~Yin, ``Unixcoder: Unified cross-modal pre-training for code representation,'' in \emph{Proceedings of the 60th Annual Meeting of the Association for Computational Linguistics (Volume 1: Long Papers)}, 2022, pp. 7212--7225.

\bibitem{shi2023cocosoda}
E.~Shi, Y.~Wang, W.~Gu, L.~Du, H.~Zhang, S.~Han, D.~Zhang, and H.~Sun, ``Cocosoda: Effective contrastive learning for code search,'' 2023.

\bibitem{ahmad2021unified}
W.~Ahmad, S.~Chakraborty, B.~Ray, and K.-W. Chang, ``Unified pre-training for program understanding and generation,'' in \emph{Proceedings of the 2021 Conference of the North American Chapter of the Association for Computational Linguistics: Human Language Technologies}, 2021, pp. 2655--2668.

\bibitem{bui2021self}
N.~D. Bui, Y.~Yu, and L.~Jiang, ``Self-supervised contrastive learning for code retrieval and summarization via semantic-preserving transformations,'' in \emph{Proceedings of the 44th International ACM SIGIR Conference on Research and Development in Information Retrieval}, 2021, pp. 511--521.

\bibitem{jain2021contrastive}
P.~Jain and A.~Jain, ``Contrastive code representation learning,'' in \emph{Proceedings of the 2021 Conference on Empirical Methods in Natural Language Processing}, 2021.

\bibitem{dong2023boosting}
Z.~Dong, Q.~Hu, Y.~Guo, Z.~Zhang, M.~Cordy, M.~Papadakis, Y.~L. Traon, and J.~Zhao, ``Boosting source code learning with data augmentation: An empirical study,'' \emph{arXiv preprint arXiv:2303.06808}, 2023.

\bibitem{van2001art}
D.~A. Van~Dyk and X.-L. Meng, ``The art of data augmentation,'' \emph{Journal of Computational and Graphical Statistics}, vol.~10, no.~1, pp. 1--50, 2001.

\bibitem{huang2021cosqa}
J.~Huang, D.~Tang, L.~Shou, M.~Gong, K.~Xu, D.~Jiang, M.~Zhou, and N.~Duan, ``Cosqa: 20,000+ web queries for code search and question answering,'' in \emph{Proceedings of the 59th Annual Meeting of the Association for Computational Linguistics and the 11th International Joint Conference on Natural Language Processing (Volume 1: Long Papers)}, 2021, pp. 5690--5700.

\bibitem{zhao2023survey}
W.~X. Zhao, K.~Zhou, J.~Li, T.~Tang, X.~Wang, Y.~Hou, Y.~Min, B.~Zhang, J.~Zhang, Z.~Dong \emph{et~al.}, ``A survey of large language models,'' \emph{arXiv preprint arXiv:2303.18223}, 2023.

\bibitem{brown2020language}
T.~Brown, B.~Mann, N.~Ryder, M.~Subbiah, J.~D. Kaplan, P.~Dhariwal, A.~Neelakantan, P.~Shyam, G.~Sastry, A.~Askell \emph{et~al.}, ``Language models are few-shot learners,'' \emph{Advances in neural information processing systems}, vol.~33, pp. 1877--1901, 2020.

\bibitem{chowdhery2022palm}
A.~Chowdhery, S.~Narang, J.~Devlin, M.~Bosma, G.~Mishra, A.~Roberts, P.~Barham, H.~W. Chung, C.~Sutton, S.~Gehrmann \emph{et~al.}, ``Palm: Scaling language modeling with pathways,'' \emph{arXiv preprint arXiv:2204.02311}, 2022.

\bibitem{touvron2023llama}
H.~Touvron, T.~Lavril, G.~Izacard, X.~Martinet, M.-A. Lachaux, T.~Lacroix, B.~Rozi{\`e}re, N.~Goyal, E.~Hambro, F.~Azhar \emph{et~al.}, ``Llama: Open and efficient foundation language models,'' \emph{arXiv preprint arXiv:2302.13971}, 2023.

\bibitem{chakraborty2022natgen}
S.~Chakraborty, T.~Ahmed, Y.~Ding, P.~T. Devanbu, and B.~Ray, ``Natgen: generative pre-training by “naturalizing” source code,'' in \emph{Proceedings of the 30th ACM Joint European Software Engineering Conference and Symposium on the Foundations of Software Engineering}, 2022, pp. 18--30.

\bibitem{mishra2022cross}
S.~Mishra, D.~Khashabi, C.~Baral, and H.~Hajishirzi, ``Cross-task generalization via natural language crowdsourcing instructions,'' in \emph{Proceedings of the 60th Annual Meeting of the Association for Computational Linguistics (Volume 1: Long Papers)}, 2022, pp. 3470--3487.

\bibitem{hu2023revisiting}
F.~Hu, Y.~Wang, L.~Du, X.~Li, H.~Zhang, S.~Han, and D.~Zhang, ``Revisiting code search in a two-stage paradigm,'' in \emph{Proceedings of the Sixteenth ACM International Conference on Web Search and Data Mining}, 2023, pp. 994--1002.

\bibitem{ChatDANCE}
Anonym, ``Anonymous replication package,'' \url{https://anonymous.4open.science/r/ChatDANCE/README.md}, 2023.

\bibitem{chatapi}
``Openai platform - chat api documentation,'' \url{https://platform.openai.com/docs/guides/chat}, 2023.

\bibitem{loshchilov2017decoupled}
I.~Loshchilov and F.~Hutter, ``Decoupled weight decay regularization,'' \emph{arXiv preprint arXiv:1711.05101}, 2017.

\bibitem{wang2020understanding}
T.~Wang and P.~Isola, ``Understanding contrastive representation learning through alignment and uniformity on the hypersphere,'' in \emph{International Conference on Machine Learning}.\hskip 1em plus 0.5em minus 0.4em\relax PMLR, 2020, pp. 9929--9939.

\bibitem{gao2021simcse}
T.~Gao, X.~Yao, and D.~Chen, ``Simcse: Simple contrastive learning of sentence embeddings,'' in \emph{2021 Conference on Empirical Methods in Natural Language Processing, EMNLP 2021}.\hskip 1em plus 0.5em minus 0.4em\relax Association for Computational Linguistics (ACL), 2021, pp. 6894--6910.

\bibitem{Meng2021coco}
Y.~Meng, C.~Xiong, P.~Bajaj, S.~Tiwary, P.~Bennett, J.~Han, and X.~Song, ``Coco-lm: Correcting and contrasting text sequences for language model pretraining,'' in \emph{35th Conference on Neural Information Processing Systems, NeurIPS 2021}.\hskip 1em plus 0.5em minus 0.4em\relax Neural information processing systems foundation, 2021, pp. 23\,102--23\,114.

\bibitem{wang2021understanding}
F.~Wang and H.~Liu, ``Understanding the behaviour of contrastive loss,'' in \emph{Proceedings of the IEEE/CVF conference on computer vision and pattern recognition}, 2021, pp. 2495--2504.

\bibitem{van2008visualizing}
L.~Van~der Maaten and G.~Hinton, ``Visualizing data using t-sne.'' \emph{Journal of machine learning research}, vol.~9, no.~11, 2008.

\bibitem{chatgpt}
\BIBentryALTinterwordspacing
Chatgpt. OpenAI. [Online]. Available: \url{https://chat.openai.com/}
\BIBentrySTDinterwordspacing

\bibitem{husain2019codesearchnet}
H.~Husain, H.-H. Wu, T.~Gazit, M.~Allamanis, and M.~Brockschmidt, ``Codesearchnet challenge: Evaluating the state of semantic code search,'' \emph{arXiv preprint arXiv:1909.09436}, 2019.

\end{thebibliography}
\end{document}